\documentclass[a4paper]{article}
\usepackage[]{amsmath}
\usepackage{amssymb}
\usepackage{amsthm} 
\usepackage{calc} 

\usepackage{layouts}

\usepackage{graphicx}
\usepackage{epstopdf}
\usepackage{enumerate}
\usepackage{mathtools}
\usepackage{lastpage}
\usepackage{lineno} 
\usepackage{pdflscape} 
\usepackage[square,comma,numbers]{natbib}
\usepackage{hyperref}
\usepackage{authblk}

\newlength{\depthofsumsign}
\allowdisplaybreaks 

\newcommand{\set}[1]{\mathcal{#1}} 	
\newcommand{\mat}[1]{\mathbf{#1}} 	

\newcommand{\transpose}[1]{#1^\intercal}

\newcommand{\der}[2]{{\delta}_{#1}^{#2}}

\begin{document}


\title{Evolutionary graph theory derived from eco-evolutionary dynamics}

\author[1]{Karan Pattni}
\author[2]{Christopher E.~Overton}
\author[3]{Kieran J.~Sharkey}
\affil[1,3]{Department of Mathematical Sciences, University of Liverpool, United Kingdom}
\affil[2]{Department of Mathematics, University of Manchester, United Kingdom}
\affil[1]{karanp@liverpool.ac.uk}
\affil[2]{c.overton@liverpool.ac.uk}
\affil[3]{kjs@liverpool.ac.uk}
\date{}
\maketitle


\begin{abstract}
A biologically motivated individual-based framework for evolution in network-structured populations is developed that can accommodate eco-evolutionary dynamics.
This framework is used to construct a network birth and death model. 
The evolutionary graph theory model, which considers evolutionary dynamics only, is derived as a special case, highlighting additional assumptions that diverge from real biological processes.
This is achieved by introducing a negative ecological feedback loop that suppresses ecological dynamics by forcing births and deaths to be coupled.
We also investigate how fitness, a measure of reproductive success used in evolutionary graph theory, is related to the life-history of individuals in terms of their birth and death rates. 
In simple networks, these ecologically motivated dynamics are used to provide new insight into the spread of adaptive mutations, both with and without clonal interference.
For example, the star network, which is known to be an amplifier of selection in evolutionary graph theory, can inhibit the spread of adaptive mutations when individuals can die naturally.
\end{abstract}

\section{Introduction}

Evolution is the process by which species adapt and change over time through the basic principles of birth, mutation, interaction and death.
It consists of ecological dynamics, which includes the change in population size and distribution, and evolutionary dynamics, which is the change in the composition of a given trait in a population. 
This process is often studied using the assumption of ecological equilibrium, i.e.~fixed population size and distribution or infinite population size.
Examples include Wright-Fisher model \cite{fisher1930,wright1949}, adaptive dynamics \citep{dieckmann1996},  evolutionary game theory \cite{maynardsmith1982,hofbauer1998,nowak2004} and evolutionary graph theory \cite{lieberman2005a}.
More recent studies consider eco-evolutionary dynamics where ecological and evolutionary dynamics interact \cite{champagnat2006,champagnat2007,constable2015,czuppon2018}, which is confirmed to be the case in real biological systems \cite{haafke2016,frickel2016}.
Our overall objective is to understand how a network-structured population affects eco-evolutionary dynamics. 
However, our primary focus here is on how ecological dynamics can be suppressed to achieve ecological equilibrium and thereby uncover the hidden ecological assumptions underpinning evolutionary graph theory.

Levins' \cite{levins1969} metapopulation model considers discrete spatial structure in the form of spatially separated sites that can be empty or occupied by a local population and whose individuals can migrate to other sites.
This model has been extended in various ways, for example, by considering a network of sites \cite{hanski2003}.
Metapopulation models are characterised by their extinction-colonisation dynamics, where local populations on occupied sites can go extinct and unoccupied sites become colonised by migrants. This means that it is possible to have both occupied and unoccupied sites.
In structured epidemic models \citep{harris1974,mollison1977}, where sites are seen as hosts who can carry infectious disease, the susceptible-infected-susceptible (SIS) dynamics consist of colonisation events in the form of susceptible hosts getting infected and extinction events in the form of infected hosts recovering.
In individual-based lattice models, such as the competing contact process \citep{durrett2009}, sites can accommodate at most one individual, so extinction on a site is a death event and colonisation is a birth event.
However, a notable exception is the individual-based framework of evolutionary graph theory \citep{lieberman2005a} where each site always has one individual present on it.
Due to this restriction, this framework differs in terms of the dynamics used in the aforementioned models where empty sites are allowed.
Dynamics that allow empty sites have been applied to biologically relevant scenarios, for example, in the case of epidemic models: foot-and-mouth disease \citep{keeling2005}, sexually transmitted diseases \citep{eames2002} and avian influenza \citep{sharkey2008}.
On the other hand, evolutionary graph theory is dominated by theoretical discussions about the importance of population structure on evolution \citep{lieberman2005a,broom2008,hadjichrysanthou2011}.
To bridge the gap between these models, we need to study them within a single framework that will allow us to view their relationship in terms of the underlying biological assumptions made at the individual level.

The modelling framework we use is Champagnat et al.'s \cite{champagnat2006} individual-based model of asexual reproduction.
Here we apply this model in the context of a network-structured population. 
We assume that individuals are distributed over a network of sites and spread by being dispersed upon birth. 
Using Champagnat et al.'s \cite{champagnat2006} model allows us to consider different evolutionary models by changing the timescale of individual-level processes.
In the limit where mutation rates tend to zero we obtain only the ecological dynamics.
As the mutation rate increases, we obtain eco-evolutionary dynamics.
In the latter case, we then consider cases where ecological and evolutionary processes happen at similar timescales which is the case in RNA viruses \citep{grenfell2004}, for example.
Examples where network-structure plays an important role include the the spread of antibiotic resistant bacteria around hospital environments \citep{lee2011}, and respiratory viruses in human contact networks \citep{salathe2010,wells2020}.

The paper is structured as follows. 
Section~\ref{sec2} describes Champagnat et al.'s \cite{champagnat2006} model and how it can be applied to allow for a network-structured population.
Section~\ref{sec:limit} gives the main result showing that ecological dynamics can be suppressed by using a negative ecological feedback loop. 
In Section \ref{sec: app 1} we construct a model with ecologically motivated dynamics, called the network birth and death model (NBD), which includes the SIS epidemic model~\citep{harris1974} as a special case. 
We then apply the result in Section~\ref{sec:limit} to the NBD model to derive evolutionary graph theory dynamics.
In Section~\ref{sec: app 2}, we investigate the long-term behaviour of the NBD model by calculating the probability of an adaptive mutant replacing a resident population both with and without clonal interference. 
We end with a brief discussion.

\section{Evolution modelling with eco-evolutionary dynamics and network structure} 
\label{sec2}
We consider a population in which individuals are distributed over a finite number of connected sites, which multiple individuals can occupy.
Individuals and sites represent different things depending on the modelling context.
Examples can be found in the metapopulation and epidemiology literature such as the fragmented habitat of fritillary butterflies \citep{hanski2017} and farms housing livestock infected with foot and mouth disease \citep{matthews2003}. 
The sites are assumed to be arranged on a network such that individuals can spread to a connected site only.
Examples of natural and artificial networks where the spread of individuals is restricted to nearest neighbours include email networks spreading computer viruses \citep{newman2002} and livestock movement networks \citep{kiss2006}.

We use Champagnat et al.'s~\cite{champagnat2006} model, an individual-level birth and death process that incorporates interaction and mutation.
In this process, the population is updated in continuous time through either a birth or death event, respectively increasing or decreasing the population size by one.
For birth events, individuals are assumed to reproduce asexually, giving rise to an offspring that is of identical type when there is no mutation or of a different type when there is mutation.
For death events, it is assumed that individuals free up any space that they previously occupied.
Deaths and births are assumed to be independent events allowing the population size to fluctuate.
Interaction between individuals can affect birth and/or death.
In particular, interaction allows the consideration of frequency dependent selection over the adaptive landscape through the use of evolutionary game theory  \cite{maynardsmith1982,ohtsuki2006,allen2017}.  
Mutation allows the introduction of a continuous number of new types into the population, allowing consideration of a richer adaptive landscape.
For example, when the evolution of a population is studied over a long period of time, multiple different types can appear that could potentially result in clonal interference \citep{gerrish1998} where two or more adaptive mutations are in competition with one another.

The mathematical description of the Champagnat et al.~\cite{champagnat2006} model is as follows. 
An individual can have $l$ real-valued phenotypic traits given by the set $\mathcal{U}\subset \mathbb{R}^l$.
The state of the population at a given point in time is given by the multiset $\set{S}$ containing the traits of each individual. 
Since $\set{S}$ is a multiset, if there are two individuals with traits $i\in\set{U}$ then there would be at least two copies of $i$ in $\set{S}$.
An individual with traits $i\in\set{U}$ is denoted $I_i$. 
The individual-level processes follow a Poisson process.
The death and birth rate of $I_i$ is respectively given by $d(i,\set{S})$ and $b(i,\set{S})$. 
The probability that an offspring of $I_i$ carries a mutation is $\mu(i)$.
Generally mutation is a fixed constant independent of phenotypic traits, though assuming there is dependence on phenotypic traits allows accounting for rare occurrences where this is the case.
For example, high antibiotic resistance in bacteria is a result of cooperative mutations where several different genes act together to provide this level of resistance \cite{martinez2000}.
In this case, the mutation probability would be higher than in the case of lower antibiotic resistance where a single gene could be acting on its own. 
The probability that $I_i$ gives birth to an offspring with trait $w$ is given by $M(i,w)$ such that $M(i,w)=0$ if $w\notin \mathcal{U}$.
Putting this together gives a model of population evolution described by a continuous time Markov process.
The infinitesimal dynamics of the state of the population $\cal{S}$ at time $t$ is described by the generator $\mathcal{L}$ that acts on real bounded functions $\phi(\set{S})$ as follows
\begin{linenomath*}
\begin{align}
	\mathcal{L}\phi(\set{S}) =& 
		\sum_{i\in \set{S}}
		[1-\mu(i)]b(i,\set{S})[\phi(\set{S}\cup\{i\})-\phi(\set{S})]
		\nonumber\\
		&+ \sum_{i\in\set{S}}
		\mu(i)b(i,\set{S})\int_{\mathbb{R}^l}[\phi(\set{S}\cup\{w\}-\phi(\set{S})]M(i,w) dw
		\nonumber\\
		&+ 
		\sum_{i\in\set{S}}d(i,\set{S})[\phi(\set{S}\setminus\{i\})-\phi(\set{S})].
\label{eq: generator}
\end{align}
\end{linenomath*}
The infinitesimal generator describes the way in which the population can change over time. 
In particular, it describes three different events that can cause the population to change.
The event described by the first line is an offspring born with no mutation, the second line is an offspring born with a mutation and the last line represents the death of an individual.
For further details on the infinitesimal generator see Appendix \ref{app: Generator Details}.

The Champagnat et al.~\cite{champagnat2006} model can be applied to a network-structured population by assuming that one of the traits is the position of the individual.
However, movement would be limited by the mutation rate $\mu$ in this case. 
To avoid this, the position of an individual is introduced as a separate characteristic. 
We then assume that individuals spread upon birth such that offspring can be placed onto a connected site where they mature immediately and remain until death.
This individual-level model will enable us to construct population-level models that use this kind of spreading mechanism between sites, such as metapopulation models \cite{levins1969}.
Examples of where this type of spreading dynamics can be used include modelling dispersal in plants \citep{fournier2004}, spread of social behaviour such as alcoholism \citep{rosenquist2010} and spread of infectious disease in epidemics.

The mathematical description of the Champagnat et al.~\cite{champagnat2006} model with network structure is as follows. 
We consider a network of $N$ distinct sites represented by a matrix $W$ with entries $W_{m,n} \geq 0$. 
Sites $m$ and $n$ are connected if $W_{m,n}>0$. 
Let $\mathcal{X}=\{1,2,\ldots,N\}$ be the set of positions individuals can occupy. 
An individual is now characterised by $i=(U_i,X_i)$ where $U_i\in \mathcal{U}$ and $X_i\in\mathcal{X}$.
This way of characterising individuals is taken from  Champagnat \& M\'el\'eard \cite{champagnat2007a} but here $\mathcal{X}$ is a discrete set.
Using the same notation as before: an individual with characteristics $i=(U_i,X_i)$ is denoted $I_i$ and the state of the population $\set{S}$ now contains elements $i$.
To represent the individuals in site $n$ we define $\set{S}_n=\{i\in \set{S}:X_i=n\}$; it  therefore follows that $\set{S}_n\subseteq \set{S}$.
We assume the network $W$ can have an impact on the death and birth rates.
The death rate of $I_i$ is given by $d(i,\set{S},W)$. 
The birth rate of $I_i$ when their offspring is spread to site $x$ is given by $b(i,x,\set{S},W)$.
Since $W$ is constant, it is dropped for brevity, i.e.~$d(i,\set{S}) = d(i,\set{S},W)$ and $b(i,x,\set{S})=b(i,x,\set{S},W)$.
The mutation functions, $\mu$ and $M$ remain the same.
The infinitesimal dynamics in this case is given by
\begin{linenomath*}
\begin{align}
	\mathcal{L}\phi(\set{S}) =& 
		\sum_{i\in \set{S}}
		\sum_{n\in\set{X}}
		[1-\mu(i)]b(i,n,\set{S})[\phi(\set{S}\cup\{(U_i,n)\})-\phi(\set{S})]
		\nonumber\\
		&+ \sum_{i\in\set{S}}
		\sum_{n\in\set{X}}
		\mu(i)b(i,n,\set{S})\int_{\mathbb{R}^l}[\phi(\set{S}\cup\{(w,n)\}-\phi(\set{S})]M(U_i,w) dw
		\nonumber\\
		&+ 
		\sum_{i\in\set{S}}d(i,\set{S})[\phi(\set{S}\setminus\{i\})-\phi(\set{S})].
\label{eq: generator}
\end{align}
\end{linenomath*}

\begin{figure}[t]
\begin{center}
\includegraphics[width=0.5\textwidth]{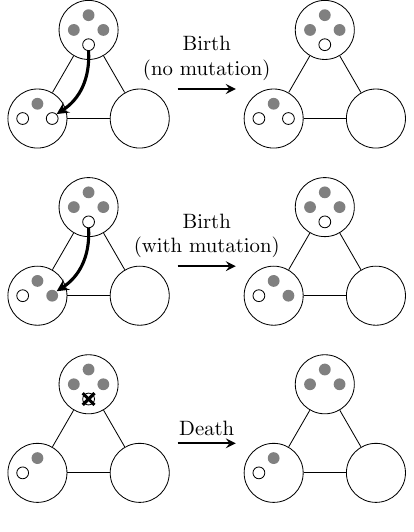}
\end{center}
\caption{\label{fig: state transitions} 
The ways in which the population state can change according to the dynamics in \mbox{Equation \eqref{eq: generator}}.  
}
\end{figure}

The dynamics in Equation \eqref{eq: generator} describe how the population state changes whenever there is birth or death in the population, this is illustrated in Figure \ref{fig: state transitions}.
In this model the size, distribution and composition of a population is given by its current state $\set{S}$.
Ecological dynamics describe how the size and distribution of a population changes due to the interaction of individuals and their environment. 
Evolutionary dynamics describe how the composition of the population changes due to evolutionary forces such as mutation and natural selection.
In the next section we show that it is possible to suppress ecological dynamics so that only the composition of the population changes when $\set{S}$ is updated.

Equation \eqref{eq: generator} will be used to calculate the probability of eventually hitting some population state $\set{A}$ from an initial state $\set{S}$, denoted $h_\set{A}(\set{S})$.
The hitting probability is found by solving
\begin{linenomath*}
\begin{align}
\begin{cases}
	\mathcal{L}h_{\set{A}}(\set{S})=0, \\
	h_{\set{A}}(\set{A})=1.
\end{cases}
\label{eq: hitting prob}
\end{align}
\end{linenomath*}
Equation \eqref{eq: generator} can also be used to find the hitting time.
The derivation of the hitting probability and time are in Appendix \ref{app: Generator Details}.


\section{Suppressing ecological dynamics in eco-evolutionary dynamics}
\label{sec:limit}
We show that we can suppress ecological dynamics in the eco-evolutionary dynamics proposed, leaving us with evolutionary dynamics that are based on ecologically motivated assumptions. 
In models that only consider evolutionary dynamics, such as the Moran process \cite{moran1960} and evolutionary graph theory \cite{lieberman2005a}, these underlying ecological assumptions are lost. 
This is because their evolutionary dynamics are directly defined from the assumption of fixed population size and distribution, rather than treating it as a consequence of suppressing ecological dynamics as is the case here.

When ecological dynamics are suppressed, the carrying capacity does not depend upon the composition of the population.
To achieve this behaviour, we create a negative ecological feedback loop that balances out opposing ecological forces pushing the system towards an equilibrium.
For example, there is negative feedback between predators and their prey where an increase in predators leads to decrease in prey and vice versa \cite{berryman2003}. 
The ecological processes that result in a birth oppose those that result in a death. 
We therefore balance out these processes such that a population converges to a given size regardless of its composition. 

The equilibrium we consider is where the population is equally distributed across all sites such that there is one individual per site giving a total population size of $N$.
There are several ways available to modify the birth and death rates so that,  whenever there is a deviation from the equilibrium, a negative ecological feedback loop comes into effect and pushes the population towards the equilibrium.
We use the Heaviside step function
\begin{linenomath*}\begin{align*}
	H_y[z]=
	\begin{cases}
		\displaystyle
		0 & z<y,\\
		1 & z \ge y,
	\end{cases}
\end{align*}\end{linenomath*}
but other functions to suppress the ecological dynamics would work equally well.
The modified death rate of $I_i$ amplifies its death rate by $c$ if present on a site with multiple occupancy but otherwise has no effect; that is, 
\begin{linenomath*}\begin{align}
	D(c,i,\set{S})=
	c^{H_2[|\set{S}_{X_i}|]}
	d(i,\set{S}) 
	\quad c\ge 1.
	\label{eq: death for constant}
\end{align}\end{linenomath*}
Similarly, the modified birth rate of $I_i$ amplifies its birth rate onto site $n$ by $c$ if site $n$ is empty but otherwise has no effect; that is,
\begin{linenomath*}\begin{align}
	B(c,i,n,\set{S})=
	c^{H_0[-|\set{S}_n|]}b(i,n,\set{S}) 
	\quad c\ge 1.
	\label{eq: birth for constant}
\end{align}\end{linenomath*}
The infinitesimal generator for the modified birth and death rates, denoted $\mathcal{L}_c$, is given by Equation \eqref{eq: generator} but with $b$ replaced by $B$ and $d$ replaced by $D$.
The parameter $c$ controls the strength of the negative ecological feedback loop's effect.
For $c=1$, there is no effect.
For $c>1$, it is more likely that individuals sharing a site will die and that offspring are placed onto empty sites. 
This effect is maximised for $c\to\infty$, which suppresses ecological dynamics resulting in fixed population size and distribution.

When ecological dynamics are suppressed, the system updates through a replacement event where a birth and a death are coupled.
This is formally shown by considering the hitting probability.
Using the generator $\mathcal{L}_c$, the hitting probability in the limit as $c \to \infty$ of the eco-evolutionary dynamics can be shown (Appendix \ref{app: Hitting probability}) to reduce to
\begin{linenomath*}\begin{align}
	h_{\set{A}}(\set{S}) =&\frac{1}{\lambda_\set{S}}\sum_{i\in \set{S}}
		\sum_{j\in\set{S}}
		r(i,j,U_i,\set{S})
		[1-\mu(i)]		
		h_{\set{A}}(\set{S}\cup\{(U_i,X_j)\}\setminus\{j\})
	\nonumber\\
	&+
	\int_{\mathbb{R}^l}
		r(i,j,w,\set{S})	
		\mu(i)		
		h_{\set{A}}(\set{S}\cup\{(w,X_j)\}\setminus\{j\})M(U_i,w) dw
\label{eq:stationary}
\end{align}\end{linenomath*}
where $\lambda_{\set{S}}$ is the rate of leaving state $\set{S}$ and $r(i,j,u,\set{S})$ is the rate at which a type $u$ offspring of $I_i$ replaces $I_j$ in state $\set{S}$.
This shows that in the limiting dynamics we have derived, the population is updated through replacement events that happen with rate $r$. 
It is shown in Appendix \ref{app: Hitting probability} that the replacement rate $r$ for Equation \eqref{eq:stationary} is given by
\begin{linenomath*}\begin{align}
r(i,j,u,\set{S})=
		b(i,X_j,\set{S})
		\frac
			{
				\displaystyle 
				d(j,\set{J})
			}
			{
				\displaystyle 
				\sum_{k\in\set{J}_{X_j}}
				\mkern-6mu 
				d(k,\set{J})
			}
		+
		d(j,\set{S})
		\frac
		{\displaystyle b(i,X_j,\set{S}\setminus\{j\})}
		{
			\displaystyle
			\sum_{k\in\set{S}\setminus\{j\}}
			b(k,X_i,\set{S}\setminus\{j\})
		}
\label{eq: rep rate}
\end{align}\end{linenomath*}
where $\set{J}=\set{S}\cup\{(u,X_j)\}$. 
We can see in $r$ that there two components, birth-death (BD), i.e.~birth followed by death, and death-birth (DB), i.e.~death followed by birth. 
The first term is a birth-death (BD) component where $I_i$ first gives birth to an offspring that is placed onto site $X_j$ who then replaces $I_j$.
The second term is a death-birth (DB) component where $I_j$ dies first and then $I_i$ gives birth to an offspring that is placed onto site $X_j$, hence replacing $I_j$.

\section{Application I: Deriving evolutionary graph theory dynamics from a network birth and death process}
\label{sec: app 1}
A model with ecologically motivated dynamics is constructed, which we will refer to as the network birth and death model (NBD). 
This model contains the SIS epidemic model~\citep{harris1974} and competing contact process~\citep{durrett2009} as special cases. 
As shown in Section~\ref{sec:limit}, we suppress ecological dynamics in NBD giving pure evolutionary dynamics based on birth and death rates. 
These evolutionary dynamics are compared to the fitness-based evolutionary graph theory dynamics to show how fitness can be interpreted in terms of birth and death rates. 
We thereby uncover hidden assumptions and provide biological insight into evolutionary graph theory dynamics.

NBD uses density-dependent regulation of population size based on Huang et al.~\cite{huang2015}. 
The intra-site competition for survival is captured through pairwise interactions. 
This competition has negative feedback such that increasing population size results in increased competition and vice versa.
For $I_i$, let $\delta_{U_i}$ be the natural death rate and $\gamma_{U_i,U_j}$ be the death rate due to competition with $I_j$. 
Huang et al.~\cite{huang2015} specifies that the inverse of $\gamma$ can be interpreted as the payoff in evolutionary game theory \cite{maynardsmith1982}.
That is, a larger payoff is received when $\gamma$ is lower.
The death rate is then given by 
\begin{linenomath*}\begin{align}
	d(i,\set{S})=\delta_{U_i}+\sum_{j\in\set{S}_{X_i}\setminus\{i\}}\gamma_{U_i,U_j}
	\label{eqDeathRate}
\end{align}\end{linenomath*}
where self-interactions have been discounted. 
It is assumed that $\gamma_{u,v}>0\ \forall u,v\in\set{U}$ to ensure negative feedback.
The birth rate is given by
\begin{linenomath*}\begin{align}
	b(i,n,\set{S})= s^{|\set{S}_n|}\beta_{U_i} W_{X_i,n}  \qquad s\in[0,1].
	\label{eqBirthRate}
\end{align}\end{linenomath*}
The birth rate of $I_i$ is $\beta_{U_i}$. 
It is weighted by $W_{X_i,n}$ to capture the network effect of its position when placing its offspring in site $n$. 
We added $s^{|\set{S}_n|}$ to capture the ability of an offspring to survive when invading site $n$ depending on its occupancy.
For $0<s<1$, there is negative feedback such that survival decreases as occupancy increases and vice versa.
For $s=0$, the convention that $0^0=1$ is used implying that there is no invasion because offspring do not survive on occupied sites.
For $s=1$, offspring always survive when invading.


NBD forms a basis for the susceptible-infected-susceptible (SIS) epidemic model \citep{harris1974} and its various extensions. 
In the SIS model, a host can be infected (I) or susceptible (S) and is represented by a node in a network. 
Infection can only spread from I to S, and is  
therefore proportional to the number of I neighbours.
Whereas I recovers (becoming S) independent of its neighbours. 
A generalisation of the SIS model that has multiple infected types is obtained from NBD as follows.
A host is represented by a site, which is S when vacant and I when occupied.
The presence of $I_i$ on a site indicates having infection $U_i$, i.e.~the phenotypic traits.
The spread of infection is represented by the birth rate in Equation \eqref{eqBirthRate}, where we set $s=0$ to restrict spread to S (vacant sites) only.
Thus, infection $U_i$ spreads with rate $\beta_{U_i}$ with weighting $W_{X_i,n}$ to capture the network effect.
Recovery from infection is represented by the death rate in Equation \eqref{eqDeathRate}, where $\gamma$ does not come in play as infection only spreads to vacant sites.
Thus, $\delta_{U_i}$ is the recovery rate from infection $U_i$.
When there are two infected types, we obtain the competing contact process \citep{durrett2009}, a model of inter-host competition between two infected types that spread using SIS dynamics.
On the other hand, we obtain Beutel et al.'s~\cite{beutel2012} model when hosts can carry multiple infections (setting $s>0$), which means there is intra-host competition.
Therefore, NBD allows us to consider a combination of inter and intra-host competition between infections.

In the framework known as evolutionary graph theory (EGT) \cite{lieberman2005a,tkadlec2019}, one considers a population of fixed size in which each vertex contains a single individual.
We wish to investigate whether EGT can be obtained from NBD. 
To do this, we first suppress ecological dynamics (see Section~\ref{sec:limit}) in NBD to obtain pure evolutionary dynamics. 
The replacement rate in this case is obtained by substituting Equations \eqref{eqDeathRate} and \eqref{eqBirthRate} into Equation \eqref{eq: rep rate}, giving
\begin{linenomath*}\begin{align}
	r(i,j,u,\set{S})=&
	s\beta_{U_i}W_{X_i,X_j}
	\frac
	{
		\delta_{U_j} + \gamma_{U_j,u}
	}
	{
		\delta_{U_j} + \gamma_{U_j,u} + \delta_{u} + \gamma_{u,U_j}
	}
	\nonumber\\
	&+
	\delta_{U_j}
	\frac
	{
		\beta_{U_i}W_{X_i,X_j}
	}
	{
		\displaystyle
		\sum_{k\in\set{S}\setminus\{j\}} \beta_{U_k}W_{X_k,X_j}
	}.
	\label{eqReplacementRate}
\end{align}\end{linenomath*}
The exponent of $s$ is 1 in the BD component as every site has one individual in this case. 
The hitting probability in NBD, denoted $h_{\set{A}}^\text{NBD}(\set{S})$, is given by substituting this replacement rate into Equation \eqref{eq:stationary}.
On the other hand, let $R(i,j,u,\set{S})$ be the replacement rate in EGT. 
Using $R$, we define an infinitesimal generator for EGT and use it to solve Equation \eqref{eq: hitting prob} to obtain the hitting probability in EGT, denoted $h_{\set{A}}^\text{EGT}(\set{S})$ (this is shown in Appendix \ref{app: Infinitesimal generator}). 
For equivalence between NBD and EGT, we check whether they have the same hitting probabilities, i.e.~$h_{\set{A}}^\text{NBD}(\set{S})=h_{\set{A}}^\text{EGT}(\set{S})$.
For the comparisons we make, we consider standard EGT definitions of the replacement rate $R$.

In EGT, three families of dynamics are generally considered~\cite{shakarian2012a}; link (L), death-birth (DB), and birth-death (BD) dynamics.
In link dynamics a link in the network is selected, then the offspring of the individual at the start of the link replaces the individual at the end of the link. In death-birth (birth-death) an individual is first selected for death (birth) before a neighbouring individual is selected for birth (death). Each of these families have two distinct cases \cite{pattni2015a}, where individuals are selected for either birth or for death.
For link dynamics we will use LB (LD) to indicate selection on birth (death). 
For BD and DB dynamics, we use an upper case letter to indicate the event in which selection happens \cite{hindersin2015}, e.g.~bD indicates selection on death.
Selection is dependent on fitness, which is the relative fecundity of individual in their competitive neighbourhood.
In evolutionary game theory, fitness is obtained by calculating the average payoff.
Payoffs depend upon the strategy played in a game specifying the rules of interactions between individuals.
We consider constant fitness, i.e.~is independent of individuals' interactions, in order to be able to make comparisons with evolutionary graph theory results based on this case, these include the circulation theorem \cite{lieberman2005a,pattni2015a} and amplification of selection \cite{lieberman2005a,hindersin2015,tkadlec2019}. 
The fitness of $I_i$ depends upon their traits only, which we denoted $F_{U_i}$.
The replacement rates for the standard EGT dynamics are given in Table \ref{tab:egtdynamics} and only hold for those states where there is 1 individual per site, i.e.~for $\set{S}$ such that $|\set{S}_x|=1 \ \forall x\in \set{X}$.

The conditions required to obtain the standard EGT dynamics from NBD are summarised in Table \ref{tab: derived standard dynamics}, excluding bD which could not be obtained (all details in Appendix \ref{app: Deriving standard}).
We note that the result for dB dynamics has previously been obtained by Maciejewski \cite{maciejewski2014} for the neutral fitness case (see also \cite{broom2010}).
The conditions specify whether $s,\beta,\delta,\gamma$ are suppressed, identical for all traits, proportional to fitness and subject to other requirements.
Due to the requirements for LD, we note that it does not extend to the variable fitness case.
The following insights are obtained from deriving the dynamics in this way:

\begin{table}
\caption{\label{tab:egtdynamics}Standard EGT dynamics. 
The shorthand notation for BD and DB dynamics follows Hindersin \& Traulsen \cite{hindersin2015} where a capital letter is used to indicate whether selection is on birth or death. 
}
\begin{tabular}{p{2cm}p{6cm}c}
Dynamics & Description & $R(i,j,u,\set{S})$\\
\hline
\\
Death-Birth-Death (Db)/ Voter Model
& 
$I_j$ dies inversely proportional to its fitness and is replaced by $I_i$ with probability proportional to $W_{X_i,X_j}$.
&
$
	\displaystyle
	\frac
		{1/F_{U_j}}
		{\displaystyle \sum_{n\in\set{S}} 1/F_{U_n}}
	\frac
		{W_{X_i,X_j}}
		{\displaystyle \sum_{k\in\set{S}\setminus\{j\}}W_{X_k,X_j}}
$
\\
Death-Birth-Birth (dB)
&
$I_j$ dies uniformly at random, i.e.~with probability $1/N$, and is then replaced by $I_i$ with probability proportional to $F_{U_i}W_{X_i,X_j}$.
&
$
	\displaystyle
	\frac
		{1}
		{N}
	\frac
		{F_{U_i}W_{X_i,X_j}}
		{\displaystyle\sum_{k\in\set{S}\setminus\{j\}}F_{U_k}W_{X_k,X_j}}
$
\\
Link-Birth (LB)
&
$I_i$ replaces $I_j$ with probability proportional to $F_{U_i}W_{X_i,X_j}$.
&
$
	\displaystyle
	\frac
		{F_{U_i}W_{X_i,X_j}}
		{\displaystyle\sum_{n,k\in\set{S}} F_{U_n}W_{X_n,X_k}}
$
\\
Link-Death (LD)
&
$I_i$ replaces $I_j$ with probability proportional to $W_{X_i,X_j}/F_{U_j}$.
&
$
	\displaystyle
	\frac
		{W_{X_i,X_j}/F_{U_j}}
		{\displaystyle\sum_{n,k\in\set{S}} W_{X_n,X_k}/F_{U_k}}
$
\\
Birth-Death-Birth (Bd)/ Invasion Process
&
$I_i$ is chosen proportional to fitness who then replaces $I_j$ with probability proportional to $W_{X_i,X_j}$. 
&
$
	\displaystyle
	\frac
		{F_{U_i}}
		{\displaystyle\sum_{n\in\set{S}} F_{U_n}}
	\frac
		{W_{X_i,X_j}}
		{\displaystyle\sum_{k\in\set{S}}W_{X_i,X_k}}
$
\\
Birth-Death-Death (bD)
&
$I_i$ is selected uniformly at random who then replaces $I_j$ proportional to $W_{X_i,X_j}/F_{U_j}$. 
&
$
	\displaystyle 
	\frac
		{1}
		{N}
	\frac
		{W_{X_i,X_j}/F_{U_j}}
		{\displaystyle\sum_{k\in\set{S}}W_{X_i,X_k}/F_{U_k}}
$
\end{tabular}
\end{table}

\begin{table}[t]
\caption{
Assumptions required for all $u,v\in\set{U}$ to obtain standard EGT dynamics from NBD. 
bD dynamics are not listed as they could not be obtained.
$\mat{1}$ is a column vector of 1s.
\label{tab: derived standard dynamics}}
\renewcommand{\arraystretch}{1.8}
\begin{tabular}{lcccc}
	Dynamics 
	& Suppressed 
	& Identical 
	& Proportional to Fitness
	& Other\\\hline
	LB 
	& $\delta_u=0$
	& $\gamma_{u,v}=\gamma_{v,u}$  
	& $\beta_u=F_u$
	& $s>0$
	\\
	\hline
	LD 
	& $\delta_u=0$
	& $\beta_{u}=\beta_{v}$  
	& $\gamma_{u,v}=1/F_u,\ u\ne v$
	& $s>0, \ |\set{U}|=2,\ \mu(i)=0$
	\\
	\hline
	Bd 
	& $\delta_u=0$
	& $\gamma_{u,v}=\gamma_{v,u}$  
	& $\beta_u=F_u$
	& $s>0$, $W\mat{1}=\mat{1}$ 
	\\
	\hline
	Db 
	& $s=0$
	& $\beta_{u}=\beta_{v}$  
	& $\delta_u=1/F_u$
	& --
	\\
	\hline	
	dB 
	& $s=0$
	& $\delta_{u}=\delta_{v}$  
	& $\beta_u=F_u$
	& --
	\\
	\hline
\end{tabular}
\end{table}

\begin{itemize}
\item Standard EGT dynamics use only one component of the replace rate in NBD (Equation \eqref{eqReplacementRate}).
Dynamics using the BD component are obtained by suppressing the natural death rate by setting $\delta_u=0\ \forall u\in\set{U}$.
This can be viewed as a biological scenario where individuals rarely die naturally but undergo intense intra-site competition with invaders. 
Individuals that can successfully invade are therefore more likely to spread.
Fitness is interpreted as the birth rate when it acts on birth.
The inverse fitness is interpreted as the death rate due to competition when it acts on death.
On the other hand, dynamics using the DB component are obtained when offspring cannot survive on occupied sites ($s=0$). 
Biologically, this can be viewed as invasion being difficult, hence individuals that can outlive their competitors are more likely to spread.
Inverse fitness is interpreted as the natural death rate when it acts on death.
Fitness is interpreted as the birth rate when it acts on birth.

\item Link dynamics is a type of BD dynamics.
In their definitions in Table \ref{tab:egtdynamics}, the order of birth and death is ambiguous and so they are classified separately from BD and DB dynamics.
This  means that intra-site competition also takes place in link dynamics.
In LB intra-site competition is independent of fitness with both individuals equally likely to die.
In LD an individual dies inversely proportional to fitness due to intra-site competition.

\item bD cannot be obtained from NBD.
It requires birth and movement to be separate, as is evident in its definition (Table \ref{tab:egtdynamics}) where the term representing birth does not specify where an offspring is placed.
In our case, birth and movement is combined as seen in Equation \eqref{eq: generator} where   $b(i,n,\set{S})$ specifies the individual that gives birth and where the offspring is placed.

\item Bd can be obtained from NBD.
Similar to bD, it is also defined with separate birth and movement (Table \ref{tab:egtdynamics}), but its movement term is independent of neighbours.
This allows us to combine movement with birth provided that it does not affect birth. 
This is only true when $W$ is right stochastic ($W\mat{1}=\mat{1}$) and, in this case, Bd and LB are equivalent \cite{pattni2015a} and therefore share the same spreading mechanism. 
This is not the case when $W$ is not right stochastic because LB allows position to affect the birth rate, which is $\beta_{U_i}\sum_{x\in\set{X}}W_{X_i,x}$ for $I_i$, whereas it is $\beta_{U_i}$ in Bd which is independent of position.

\item Db and dB can be obtained from SIS-type epidemic dynamics.
This is because they share the same spreading mechanism and do not have intra-site competition.
Using the multi-strain SIS model described above, they are obtained by letting $\beta_u\to\infty\ \forall u\in\set{U}$, resulting in vacant sites immediately being occupied by offspring.
This is how Durrett \& Levin \cite{durrett1996} use the contact process to obtain the voter model \citep{holley1975}, which has identical dynamics to Db. 
This illustrates that pathogen evolution, at least at the between host level, is likely to behave similarly to death-birth evolutionary dynamics rather than birth-death. 

\item When Db and dB are derived from NBD there is no self-replacement, i.e.~individuals cannot be replaced by their own offspring.  
The DB component of the replacement rate in NBD (Equation\eqref{eqReplacementRate}) specifies that death happens first followed by birth, therefore preventing self-replacement.
The standard definitions (Table \ref{tab:egtdynamics}) allow self-replacement for both BD and DB type dynamics, but for dynamics derived from NBD this is only possible for BD type dynamics.
Note that Db can be obtained from our derivation of LD when $W$ is left stochastic, i.e. $\transpose{W}\mat{1}=\mat{1}$.
In this case self-replacement is allowed as LD is a type of BD dynamics.
However, using this definition is limited due to the restrictions on LD (see Table \ref{tab: derived standard dynamics}).

\end{itemize}

Other non-standard EGT definitions of the replacement rate $R$ can also be considered. 
For example, Zukewich et al.~\cite{zukewich2013} combines Bd and Db using a parameter to allow a smooth transition between the two.
Setting the parameter to 1 gives Bd, 0 gives Db and a value in the 0 to 1 range gives a combination of them.
The replacement rate in NBD, Equation \eqref{eqReplacementRate}, provides a biologically motivated alternative to constructing hybrid models comprising both Bd and Db dynamics.
Another example is Kaveh et al.'s \cite{kaveh2015} DB dynamics that are not based on fitness and are obtained from Equation \eqref{eqReplacementRate} by setting $s=0$. 
They use parameters similar to $\delta$ and $\beta$ that give the likelihood of birth and death. 
However, their parameters are weights rather than rates since their system is constructed in discrete time. 


\section{Application II: Long-term behaviour of a network birth and death process}
\label{sec: app 2}
The long-term behaviour of NBD is analysed in the cases with and without clonal interference.

\subsection{No clonal interference}
It is assumed that adaptive mutations arise in succession as in Muller's \cite{muller1932} classical model.
Here, a resident population can only be invaded by one type of mutant at a time so that there is no interference from other mutations.
This means that either the current resident or mutant goes extinct before another mutation arises.
This behaviour is obtained in the rare mutation limit, $\mu(i)\to 0\ \forall i$, in Champagnat et al.'s \cite{champagnat2006} model, who derive adaptive dynamics \citep{metz1995,dieckmann1996} in this limit.
We assume no mutation, $\mu(i)=0\ \forall i$, as the results are identical for the evolutionary scenario considered.


We consider an evolutionary scenario that plays out between two types; resident (trait 0) and mutant (trait 1), so $\set{U}=\{0,1\}$.
A mutant is introduced into a resident population by replacing a resident selected uniformly at random, and the individuals then compete with each other. In the limit of infinite time, the population will go to extinction since this is the only true absorbing state. However, before this the population must reach a state where only one type remains. We are interested in the hitting probability for the set of states where only the mutant type remains, since this provides a measure of how successful the mutant type is.

The probability of reaching a state with only the mutant type is formally defined as follows. 
Let $\set{R}=\{\set{S}:U_i=0\ \forall i\in\set{S}\}$ be the set of states where only the resident type remains; i.e.~there is at least one resident but no mutants.
Similarly, let $\set{M}=\{\set{S}:U_i=1\ \forall i\in\set{S}\}$ be the set of states where only the mutant type remains. 
We want the probability of hitting $\set{M}$ from an initial population state $\set{S}$.
This probability, denoted $\rho(\set{S})$, is calculated by solving the Equation
\begin{linenomath*}
\begin{align}
	\begin{cases}
		\mathcal{L}_c\rho(\set{S})=0\quad &\set{S}\notin \set{M}\cup\set{R},\\
		\rho(\set{S})=0 \quad &\set{S}\in \set{R},\\
		\rho(\set{S})=1\quad &\set{S} \in \set{M}.
	\end{cases}
\label{eq: fixation}
\end{align}
\end{linenomath*}
The first line says the population is in a state with both mutants and residents so the infinitesimal generator for modified dynamics, $\mathcal{L}_c$, is used to specify how the hitting probability changes with an infinitesimally small change in time.
Recall that $c$ controls the strength of the negative ecological feedback loop.
The second line says the hitting probability is 0 as the population cannot hit a state with only mutants starting from a state with only residents.
The third line says the hitting probability is 1 as the population is already in a state with only mutants.

Since there is no mutation, a mutant is assumed to be equally likely to appear on any given site.
We will therefore consider an initial state with 1 mutant and $N-1$ residents where each site is occupied by one individual only.
This allows us to make comparisons with evolutionary graph theory where these are the only possible initial states.
The average of $\rho$
 for an initial mutant placed uniformly at random is given by
\begin{linenomath*}
\begin{align}
	\bar{\rho}=\frac{1}{N}\sum_{n=1}^N \rho(\{\set{S}_0\setminus\{(0,n)\}\}\cup\{(1,n)\}),
	\label{eq: avg fixation}
\end{align}	
\end{linenomath*}
where $\set{S}_0=\{(0,1),(0,2),\ldots,(0,N)\}$ is the state with one resident on each site.
We will use $\bar{\rho}$ to compare different NBD dynamics when mutants are assumed to have identical death rates to the residents ($\delta_0=\delta_1$ and $\gamma_{u,v}=\gamma_{v,u}\ \forall u,v\in\{0,1\}$) but an advantageous birth rate ($\beta_1>\beta_0$). 
The cases of the dynamics we compare are given in Table \ref{tab: dynamics compared}. 

\begin{table}[t]
\caption{
The cases of the dynamics compared in the NBD model.
For all cases mutants are assumed to have identical death rates to the residents ($\delta_0=\delta_1$ and $\gamma_{u,v}=\gamma_{v,u}\ \forall u,v\in\{0,1\}$) but an advantageous birth rate ($\beta_1>\beta_0$).  
\label{tab: dynamics compared}}
\renewcommand{\arraystretch}{1.8}
\begin{tabular}{cll}
	Case 
	& Dynamics
	& Parameter Values
	\\\hline
	(i)
	& SIS 
	& $\delta_0=\delta_1>0$ and $s=0$
	\\
	(ii)
	& SIS with invasion
	& $\delta_0=\delta_1>0$ and $s=1$
	\\
	(iii)
	&SIS with invasion and no natural death 
	& $\delta_0=\delta_1=0$ and $s=1$
\end{tabular}
\end{table}	 


\begin{figure}[t]
\begin{center}
\includegraphics[width=\textwidth]{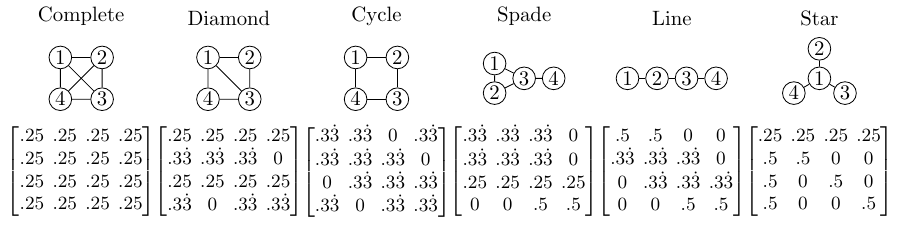}
\end{center}
\caption{\label{fig: networks} 
4-node networks where all outgoing edges from a site with $k$ neighbours are weighted $1/(k+1)$ including self-loops (not depicted on the networks).  
}
\end{figure}

\begin{figure}[t]
\begin{center}
\includegraphics[width=0.6\textwidth,trim={1cm 1cm 1cm 1cm},clip]{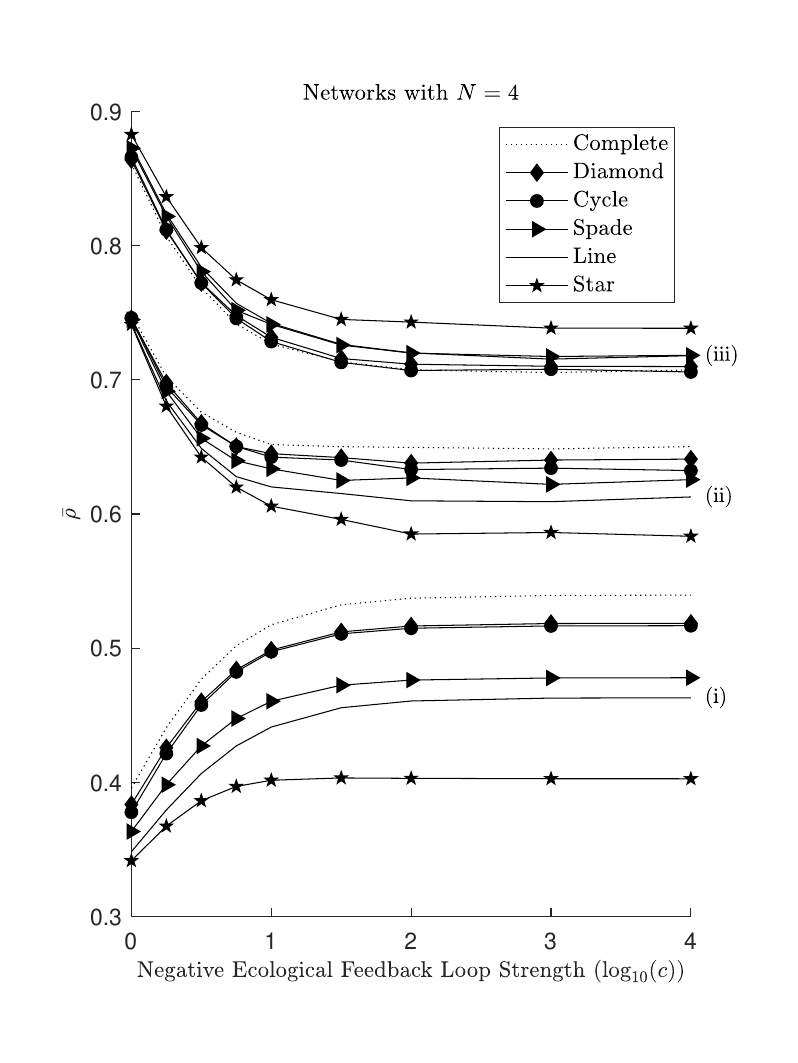}
\end{center}
\caption{\label{fig: fixation with natural death} 
$\bar{\rho}$ in the networks given in Figure \ref{fig: networks} for the three cases in Table \ref{tab: dynamics compared}, they are labelled (i)-(iii) on the right-hand side.
For all cases, we set $\beta_0=3,\ \beta_1=10,$ and $\gamma_{u,v}=5\ \forall u,v\in\{0,1\}$.
For cases (i) and (ii) we have $\delta_{u}=1$.   
In case (i), $\bar{\rho}$ is calculated by analytically solving Equation \eqref{eq: fixation}, for details on how to do this using a state transition matrix see Hindersin and Traulsen \cite{hindersin2015}. 
In cases (ii) and (iii), $\bar{\rho}$ are calculated by running $10^5$ simulations (see Appendix \ref{app: Simulation Details} for details).
}
\end{figure}

In Figure \ref{fig: fixation with natural death}, $\bar{\rho}$ is plotted against the negative feedback amplifier $c$ for the networks shown in Figure \ref{fig: networks}.
It is observed that 
\begin{linenomath*}
\begin{align}
	\bar{\rho}^\text{(i)} < \bar{\rho}^\text{(ii)} < \bar{\rho}^\text{(iii)}
	\label{eq: fixation order}
\end{align}
\end{linenomath*}
for all networks where $\bar{\rho}^\text{(i)},\ \bar{\rho}^\text{(ii)}$ and $\bar{\rho}^\text{(iii)}$  are values of $\bar{\rho}$ for cases (i), (ii) and (iii) in Table \ref{tab: dynamics compared} respectively.
SIS-type dynamics are therefore the least beneficial for the spread of an advantageous mutant.
Moving from (i) to (ii), shows that allowing invasion is beneficial, since $\bar{\rho}$ shifts up with the networks maintaining their order.
As we move from (ii) to (iii), disallowing natural death provides a further benefit, since $\bar{\rho}$ shifts higher up.
However, the networks now change their order. 
In particular, the combined effect of allowing invasion and disallowing natural death is largest in the star network and smallest in the complete network. 

To investigate the difference between the complete and star networks, we analytically calculate $\bar{\rho}$ when ecological dynamics are suppressed ($c\to\infty$).
In this case the population cannot go extinct and therefore fixates in $\set{M}$ or $\set{R}$, i.e.~indefinitely remains in a state where there is only one type.
Therefore, $\bar{\rho}$ is called the average fixation probability of mutants, a quantity widely studied in population evolution \cite{patwa2008}.

\subsubsection{Average hitting probability for a complete network}
Consider a complete network with arbitrary weights 
\begin{linenomath*}
\begin{align}
	W_{i,j}=w>0\qquad \forall i,j\in\{1,\ldots,N\}.
\label{eq: complete weights}
\end{align}
\end{linenomath*}
The average fixation probability of a single initial mutant in this network is given by the formula of Karlin and Taylor \cite{karlin1975} 
\begin{linenomath*}\begin{align}
	\bar{\rho}_{\text{comp}}=
		\left(
			1 + \sum_{j=1}^{N-1}\prod_{k=1}^j 
			r_k
	    \right)^{-1}
\label{eq: k and t}
\end{align}\end{linenomath*}
where $r_k$ for NBD is given by
\begin{linenomath*}\begin{align}
	r_k &= 
	\frac{
		s\beta_{0} w
		\frac{
			(\delta_1 + \gamma_{1,0})
		}{
			(\der{0}{} + \der{1}{}+ \gamma_{0,1}{}  + \gamma_{1,0}{})
		}
		+
		\der{1}{}
		\frac{
			\beta_{0} 
		}{
			(N-k) \beta_{0}  + (k-1) \beta_{1} 
		}
	}
	{
		s\beta_{1} w
		\frac{
			(\der{0}{} + \gamma_{0,1}{})
		}{
			(\der{0}{}  + \der{1}{}+ \gamma_{0,1}{} + \gamma_{1,0}{})
		}
		+
		\der{0}{}
		\frac{
			\beta_{1} 
		}{
			(N-k-1)\beta_{0} + k\beta_{1} 
		}
	}.
	\label{eq: bias rk}
\end{align}\end{linenomath*}
The term $r_k$ is the backward bias of mutants or forward bias of residents. 
It is obtained by dividing the rate of a resident increasing by the rate of a mutant increasing in a state where there are $k$ mutants (and $N-k$ residents). Details given in Appendix \ref{app: Derivation of bias}.

Table \ref{tab: dynamics combs} shows the bias and average fixation probability for cases (i), (ii), (iii) in Table \ref{tab: dynamics compared}. 
The probabilities shown are a closed-form version of Equation \eqref{eq: k and t} such that: 
(i) is obtained from Hindersin and Traulsen \cite{hindersin2015}; (ii) is not shown as there is no simple analytical form; (iii) is obtained from the Moran probability \citep{moran1959} as the bias is constant.
Analysis of the biases in Table \ref{tab: dynamics combs} reveals
\begin{linenomath*}
\begin{align}
	r_k^\text{(i)} > r_k^\text{(ii)} > r_k^\text{(iii)}, 
	\label{eq:bias order}
\end{align}
\end{linenomath*}
where $r_k^\text{(i)},r_k^\text{(ii)},r_k^\text{(iii)}$ is the bias in cases (i), (ii), (iii) respectively.
The proof is given in Appendix \ref{app: Showing strict}. 
The key requirement for Equation \eqref{eq:bias order} to hold is $\beta_1>\beta_0$, which we assume is true. 
Equation \eqref{eq:bias order} implies that Equation \eqref{eq: fixation order} holds for all $N>1$ because, as seen in Equation \eqref{eq: k and t}, a larger bias gives a lower fixation probability.
The difference between these cases diminishes as $N\to\infty$ because their biases all converge to $\beta_0/\beta_1$, this is seen in Table \ref{tab: dynamics combs} where $\displaystyle \lim_{N\to\infty}r_{k}=\beta_0/\beta_1\ \forall k$ in all cases.  	

\begin{figure}[t]
\includegraphics[width=0.5\textwidth]{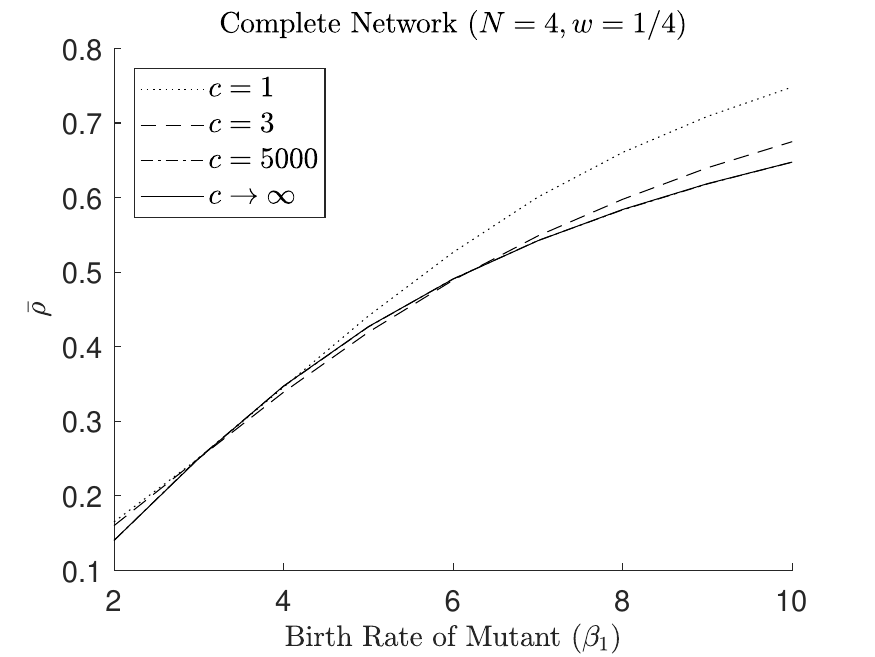}
\includegraphics[width=0.5\textwidth]{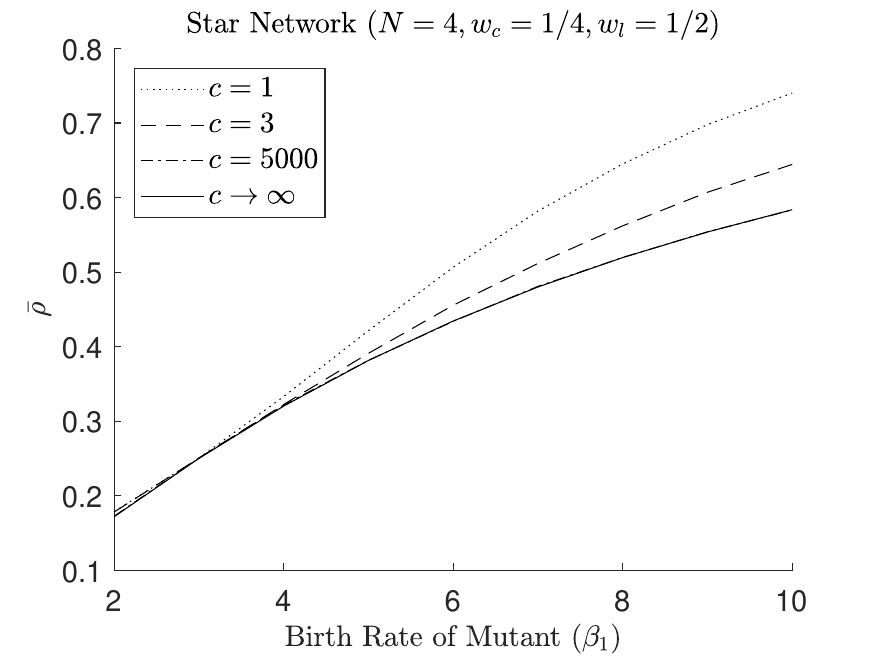}
\caption{\label{fig: convergence for comp} 
$\bar{\rho}$ plotted against the birth rate of a mutant ($\beta_1$) in the complete and star networks when using NBD dynamics with $\beta_0=3,\ \delta_u=1$ and $\gamma_{u,v}=5\ \forall u,v\in\{0,1\}$.
For $c\to\infty$, $\bar{\rho}$ is given by $\bar{\rho}_\text{comp}$ for the complete network and $\bar{\rho}_\text{star}$ for the star network.
For other values of $c$, $\bar{\rho}$ is calculated by running $10^6$ simulations (see Appendix \ref{app: Simulation Details} for details).
As $c$ gets larger we can see that $\bar{\rho}$ converges to $\bar{\rho}_\text{comp}$ for the complete network and $\bar{\rho}_\text{star}$ for the star network. 
In particular, the plots for $c=5000$ and $c\to \infty$ overlap with one another.
}
\end{figure}

Figure \ref{fig: convergence for comp} shows $\bar{\rho}$ (Equation \ref{eq: avg fixation}) plotted against $c$ in the complete network. 
Numerically, we observe that $\bar{\rho}$ converges to $\bar{\rho}_\text{comp}$ as $c$ gets large, showing that the negative ecological feedback loop functions as desired. 

\begin{table}[t]
\caption{
Bias and average fixation probability in complete network for NBD. 
Cases considered assume an advantageous mutant with $\beta_1>\beta_0$.
\label{tab: dynamics combs}}
\renewcommand{\arraystretch}{2.2}
\begin{tabular}{p{4cm}ll}
	Case/ EGT Dynamics
	& Bias $\left(r_k,\ k=1,\ldots,N-1\right)$
	& Avg.~Fixation Prob.~($\bar{\rho}_\text{comp}$)\\\hline
	(i) SIS-type dynamics ($\delta_{0}=\delta_{1}>0$, $s=0$)/ dB
	& $\displaystyle \frac{\beta_{0}}{\beta_{1}}\frac{k\beta_{1}+(N-k-1)\beta_{0}}{(N-k)\beta_{0}+(k-1)\beta_{1}}$
	& $\displaystyle \frac{N-k}{N}\frac{1-(\beta_{0}/\beta_{1})}{1-(\beta_{0}/\beta_{1})^{N-1}}$
	\\\hline
	(ii) then allow invasion ($\delta_0=\delta_1=\delta>0,\ s=1$)/ None
	& $\displaystyle\frac
		{s\beta_0w\frac{1}{2}+\delta\frac{\beta_0}{(N-k)\beta_0+(k-1)\beta_1}}
		{s\beta_1w\frac{1}{2}+\delta\frac{\beta_1}{(N-k-1)\beta_0+k\beta_1}}$
	& No simple analytical form.
	\\\hline
	(iii) then disallow natural death ($\delta_0=\delta_1=0,\ s=1$)/ LB, Bd 
	& ${\beta_{0}}/{\beta_{1}}$
	& $\displaystyle \frac{1-(\beta_{0}/\beta_{1})}{1-(\beta_{0}/\beta_{1})^N}$
	\\\hline
\end{tabular}
\end{table}

\subsubsection{Average hitting probability for a star network}\label{sec512}
Consider the star network with weights 
\begin{linenomath*}
\begin{align}
    W_{i,j}=
    \begin{cases}
	    w_c>0 & i=1 \text{ and } j=1,2,\ldots,N, \\
	    w_l>0 & i=2,3,\ldots,N \text{ and } j = 1, \\
	    0 & \text{otherwise}.
	\end{cases}
	\label{eq: star network}
\end{align}
\end{linenomath*}
Site 1 is connected to all other $N-1$ sites, which are only connected to site 1.
We call the individual in site 1 the centre and individuals in all other sites leaves.

Let there be $k$ mutant ($N-1-k$ resident) leaves.
When ecological dynamics are suppressed in NBD, a mutant centre replaces a resident leaf with rate
\begin{linenomath*}\begin{align}
		(N-1-k)
		\left(
		sw_c\beta_1
		\frac{
			\delta_{0}+\gamma_{0,1}
		}{
			\delta_0+\delta_1+\gamma_{0,1}+\gamma_{1,0}
		}
	+
		\delta_{0}
		\right),
	\label{eq: leaf replaced}
\end{align}\end{linenomath*}
whereas a resident centre is replaced by a mutant leaf with rate
\begin{linenomath*}\begin{align}
		k
		\left(
		sw_l\beta_{1}
		\frac{
			\delta_0+ \gamma_{0,1}
		}{
			\delta_0+\delta_1+\gamma_{0,1}+\gamma_{1,0}
		}
	+
		\delta_0
		\frac{
			\beta_{1}
		}{
			(N-1-k)\beta_0 + k\beta_{1}
		} 
		\right).
	\label{eq: centre replaced}
\end{align}\end{linenomath*}
Using these rates, the average fixation probability in the star network, denoted $\bar{\rho}_\text{star}$, is calculated using Hadjichrysanthou et al.'s~\cite{hadjichrysanthou2011} formula (see Appendix \ref{app: Average fixation}).
In Appendix \ref{app: Proof for star}, we show that that Equation \eqref{eq: fixation order} holds when $N\to\infty$. 
Note that for the complete network we were able to show this for all $N$.

Equations \eqref{eq: leaf replaced} and \eqref{eq: centre replaced} reveal the interplay between the BD and DB components.
In particular, consider the case where $w_c=1/N$ and $w_l=1/2$ so that the birth rate is exactly $\beta_u\ \forall u\in\{0,1\}$. 
As $N$ gets larger, the BD component in Equation \eqref{eq: leaf replaced} and the DB component in Equation \eqref{eq: centre replaced} get smaller.
This means that the highly connected centre is more reliant on DB to spread its offspring whereas the less connected leaves are more reliant on BD to spread their offspring.

Figure \ref{fig: convergence for comp} shows $\bar{\rho}$ (Equation \ref{eq: avg fixation}) for different values of $c$ in the star network. 
Its qualitative properties are similar to that of the complete network and we once again see that the negative ecological feedback loop functions as desired. 
%

\subsubsection{Comparison of average hitting probabilities for complete and star networks}
Lieberman et al.~\cite{lieberman2005a} show that when using Bd dynamics, the star network amplifies the average fixation probability when compared to the complete network, i.e.~$\bar{\rho}_\text{star}>\bar{\rho}_\text{comp}$.  
By using NBD we can gain further insight as to why this is the case.

\begin{figure}[t]
\begin{center}
\includegraphics[width=0.6\textwidth]{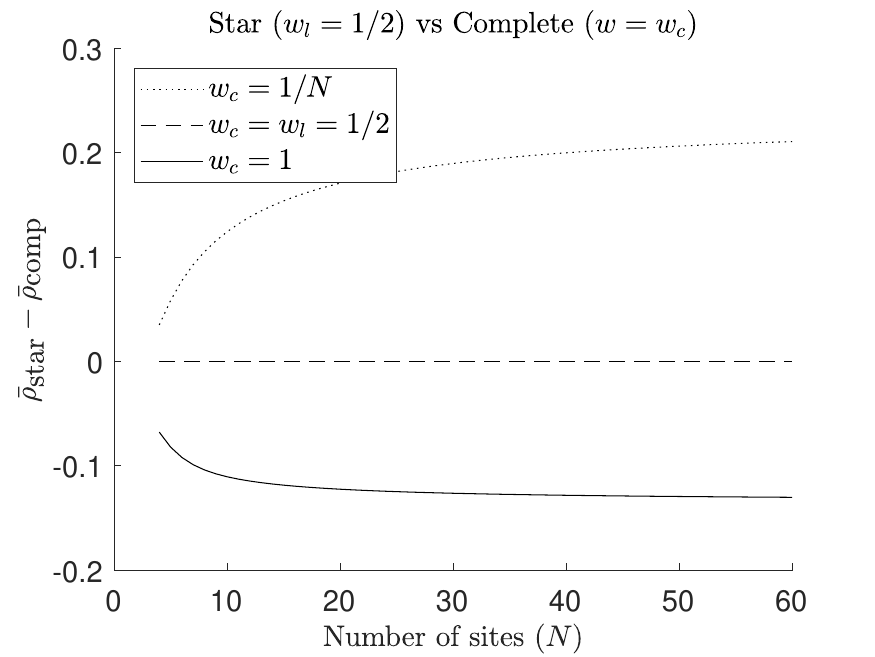}
\end{center}
\caption{\label{fig: source sink} 
Plot of $\bar{\rho}_\text{star}-\bar{\rho}_\text{comp}$ against the number of sites ($N$) when using NBD (with suppressed ecological dynamics) with $s=1$, $\beta_0=3,\ \beta_1=8,\ \gamma_{u,v}=5$ and $\delta_u=0\ \forall u,v\in\{0,1\}$.
It shows that the star network is no longer an amplifier when $w_c>w_l$.
}
\end{figure}

In source-sink metapopulation dynamics \citep{pulliam1988}, a source is a site that is a net exporter of individuals whereas a sink is a site that is a net importer of individuals.
A source site is advantageous in comparison to a sink site as more offspring are produced.
In the star network, to check whether a leaf or the centre behaves as a source we consider the case (iii) (from Table \ref{tab: dynamics compared}) but with neutral residents and mutants i.e.~$\beta_u=\beta$, $\delta_u=\delta$ and $\gamma_{u,v}=\gamma$ $\forall\ u,v\in\{0,1\}$. 
From Equation \eqref{eq: leaf replaced}, the rate at which a leaf is replaced by on offspring of the centre is 
\begin{linenomath*}
\begin{align*}
	w_c\beta\frac{1}{2},
\end{align*}
\end{linenomath*}
whereas, from equation \eqref{eq: centre replaced}, the centre is replaced by the offspring of a leaf is
\begin{linenomath*}
\begin{align*}
	w_l\beta\frac{1}{2}.
\end{align*} 
\end{linenomath*}
A leaf is therefore a source when $w_l>w_c$ and sink when $w_l<w_c$.
When calculating $\bar{\rho}_\text{star}$ a randomly placed initial mutant is more likely to be a leaf.
An advantageous mutant is therefore does better when leaves are sources.
In particular, for Bd dynamics leaves are sources as $w_c=1/N$ and $w_l=1/2$ so the star network amplifies the average fixation probability of an advantageous mutant.
This is verified in Figure \ref{fig: source sink} which illustrates that $\bar{\rho}_\text{star}>\bar{\rho}_\text{comp}$ when $w_l>w_c$, $\bar{\rho}_\text{star}<\bar{\rho}_\text{comp}$ when $w_l<w_c$, and $w_l=w_c$ is the boundary between amplification and suppression where $\bar{\rho}_\text{star}=\bar{\rho}_\text{comp}$.

\begin{figure}[t]
\begin{center}
\includegraphics[width=0.6\textwidth]{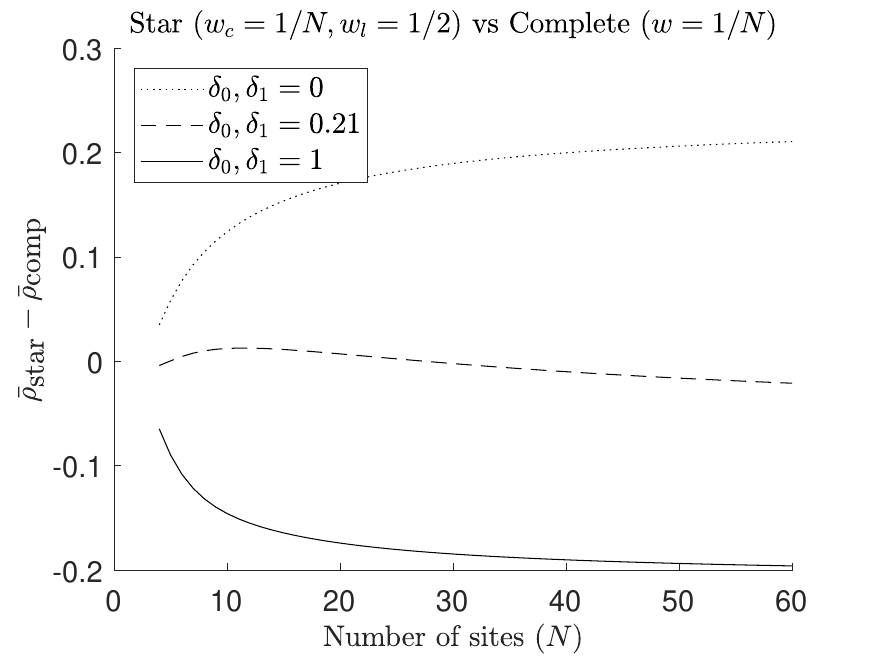}
\end{center}
\caption{\label{fig: comparison analytical} 
Plot of $\bar{\rho}_\text{star}-\bar{\rho}_\text{comp}$ against the number of sites ($N$) when using NBD (with suppressed ecological dynamics) with $s=1$, $\beta_0=3,\ \beta_1=8,\ \gamma_{u,v}=5$ and $\delta_u=0,0.21,1\ \forall u,v\in\{0,1\}$.
It shows that the star network is no longer an amplifier as the natural death rate increases.
}
\end{figure}

The natural death rate plays a fundamental role since it can prevent a leaf from being a source.
This is seen in case (ii) (from Table \ref{tab: dynamics compared}) when comparing the centre to a leaf when both residents and mutants are neutral.
That is, from Equation \eqref{eq: leaf replaced}, the rate at which a leaf is replaced by an offspring of the centre is 
\begin{linenomath*}
\begin{align*}
	w_c\beta\frac{1}{2} +\delta,
\end{align*}
\end{linenomath*}
whereas, from Equation \eqref{eq: centre replaced}, the rate at which the centre is replaced by an offspring of a leaf is
\begin{linenomath*}
\begin{align*}
	w_l\beta\frac{1}{2} + \frac{\delta}{N}.
\end{align*}
\end{linenomath*}
The natural death rate can therefore prevent a leaf from being a source when $w_l>w_c$. 
In particular, when the centre dies, leaves compete with one another for their offspring to be the replacement but, when a leaf dies, an offspring of the centre is the only replacement.
Another way to look at this is that a natural death rate limits the amount of time a leaf has to spread its offspring before it dies.
The natural death rate can therefore suppress the fixation probability of an advantageous mutant in the star network.
This is verified in Figure \ref{fig: comparison analytical}, where increasing the death rate causes $\bar{\rho}_\text{star}-\bar{\rho}_\text{comp}$ to decrease, such that the star network is no longer an amplifier of selection.
This is consistent with Hadjichrysanthou et al.~\cite{hadjichrysanthou2011} which shows that the star is not an amplifier under Db and dB dynamics.

\subsection{With clonal interference}
Here we no longer assume that adaptations are successive, and instead take into account the effect of clonal interference, which has been demonstrated in a range of asexual organisms \citep{kao2008}.
For clonal interference in unstructured populations, it has been shown that the fixation probability of a beneficial mutation decreases as the population size and mutation rate increases \citep{gerrish1998}.
The inclusion of clonal interference will therefore provide a better understanding of the impact that population structure has on the success of an adaptive mutation.

To study the effect of clonal interference we consider the evolutionary scenario considered in Gerrish \& Lenski \cite{gerrish1998}. 
Here, a resident population (type $0$) can be invaded by two kinds of mutant, an original mutant (type $1$) and a superior mutant (type $2$), i.e.~$\set{U}=\{0,1,2\}$. 
The population initially consists of the resident and original mutant types such that, to be consistent with the no clonal interference case, an original mutant is introduced into a resident population by randomly replacing a resident. A superior mutant is introduced later into the population through random mutation in the resident type.
Therefore, there is initially competition between the resident and original mutant types, but the superior mutant type can interfere.
We are interested in the probability of reaching a state where only the original mutant type remains since it is a measure of its success in the presence of clonal interference.

To define this formally, we assume that a resident has constant mutation probability and that its mutated offspring is a superior mutant, i.e. $\mu(i)=\mu$ if $U_i=0$ but $\mu(i)=0$ otherwise, and $M(u,v)=1$ if $u=0,v=2$ but $M(u,v)=0$ otherwise.
Note that the integral in Equation \eqref{eq: generator} is changed to a summation because of the discrete number of mutations.
Let $\set{R}$ be the set of states where only the resident type remains as we previously defined, and $\set{M}_1$ and $\set{M}_2$ be the set of states with all type 1 and type 2 individuals respectively.
We want to calculate the probability, $\psi$, of hitting $\set{M}_1$ conditional on not hitting $\set{R}\cup\set{M}_2$ starting from an initial state $\set{S}$.
When using the modified dynamics, this is found by solving the Equation
\begin{linenomath*}
\begin{align}
	\begin{cases}
		\mathcal{L}_c\psi(\set{S})=0\qquad &\set{S}= \{(u,x)\notin\set{M}_1:\exists u=1\}\\
		\psi(\set{S})=0 \quad &\set{S}\in \set{R}\cup\set{M}_2,\\
		\psi(\set{S})=1 \quad &\set{S}\in \set{M}_1.
	\end{cases}
	\label{eq: hitting with interference}
\end{align} 
\end{linenomath*}
We consider the initial states with 1 original mutant and $N-1$ residents where each site is occupied by one individual only, so the average of $\psi$ for a randomly placed initial original mutant is   
\begin{linenomath*}
\begin{align*}
	\bar{\psi}=\frac{1}{N}\sum_{i\in\set{S}_0} \psi(\{\set{S}_0\setminus\{i\}\}\cup\{(1,X_i)\}).
\end{align*}	
\end{linenomath*}
Note that when $\mu=0$, there is no clonal interference and we have that $\psi=\rho$ (so $\bar{\psi}=\bar{\rho}$).

\begin{figure}[t]
\begin{center}
\includegraphics[width=0.6\textwidth]{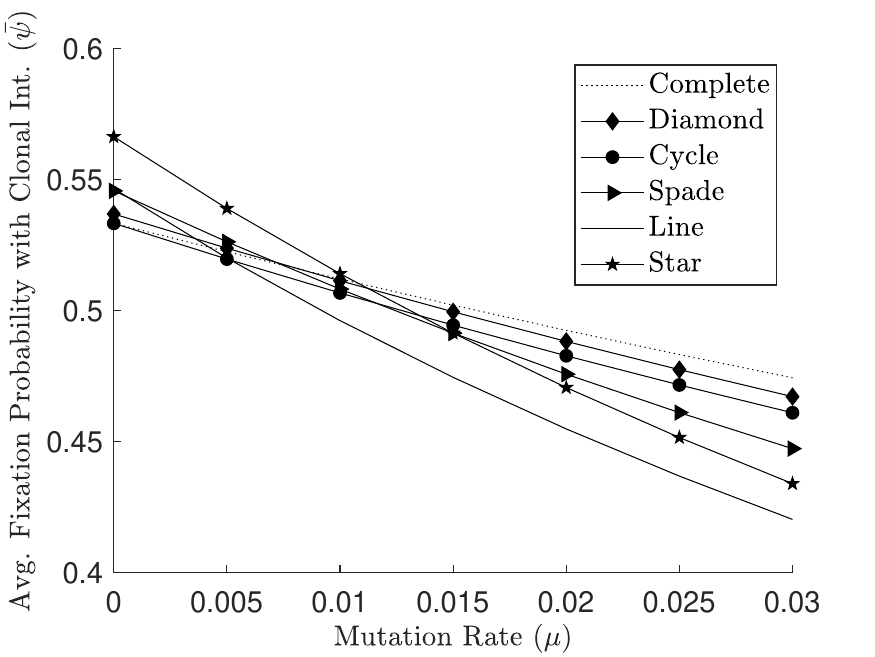}
\end{center}
\caption{\label{fig: fixation before CI}
Comparison of average fixation probability with clonal interference ($\bar{\psi}$) for 4-node networks in NBD (with ecological dynamics suppressed). 
The parameters are set as follows $s=1$, $\beta_0=1$, $\beta_1=2$, $\beta_2=3$, $\delta_u=0$, $\gamma_{u,v}=1$ for all $u,v\in\{0,1,2\}$.
$\bar{\psi}$ is calculated by analytically solving Equation \eqref{eq: hitting with interference} using a state transition matrix \cite{hindersin2015}.
} 
\end{figure}

Clonal interference reduces the amount of time that a mutant of type $1$ has to fixate, since the longer it takes the more likely a mutant of type 2 will appear. 
Without clonal interference, the complete network has the lowest fixation time whereas, for example, the star network is substantially higher \citep{frean2013,tkadlec2019,moller2019}.
We should therefore expect the complete network to be least affected as the mutation rate increases in comparison to the star and other networks.
To show that this is indeed the case, we plot $\bar{\psi}$ for different mutation rates in Figure \ref{fig: fixation before CI} for the networks with four sites given in Figure \ref{fig: networks} when ecological dynamics are suppressed in NBD ($c\to\infty$).
The population does not go extinct in this case and fixates in either $\set{M}_1$ or $\set{M}_2$, so $\bar{\psi}$ is the average fixation probability of an original mutant with clonal interference.
The average fixation probability under clonal interference decreases in all networks, with the complete network being the least affected since it has the lowest fixation time.

The circulation theorem \citep{maruyama1974,lieberman2005a} in EGT identifies networks whose fixation probability is equal to the Moran probability.
This theorem holds for simple evolutionary dynamics \cite{pattni2015a}, and generally fails for other dynamics.
Here we see it failing due to clonal interference. 
In Figure \ref{fig: fixation before CI}, we see that the fixation probability is identical for complete and circle networks because the circulation theorem holds as $\mu=0$, but this is no longer the case when $\mu>0$.

\section{Discussion}
We have reinterpreted Champagnat et al.'s \cite{champagnat2006} model to enable eco-evolutionary dynamics in a network-structured population.
This model is based on individual-level ecological processes that allow the population size, distribution and composition to change. 
It represents an advance on current evolutionary graph theory models which only allow the composition of the population to change.
We can therefore consider cases with more complex dynamics, such as pathogen evolution \cite{grenfell2004}.
In Sections \ref{sec:limit} and \ref{sec: app 1} we showed, using a negative ecological feedback loop, that suppression of ecological dynamics leaves the pure evolutionary dynamics of evolutionary graph theory \cite{lieberman2005a}. 
However, this process highlights the extreme assumptions required and the departure from the underpinning ecological processes.
In particular, the fixed population size and distribution obtained by suppressing ecological dynamics are an exception  \citep{melbinger2010,cremer2011} and can prevent us capturing biological processes accurately.
It is therefore useful to know how results will change when moving away from the extreme assumptions in evolutionary graph theory.
We showed that there is a weakened effect of network structure (Figure \ref{fig: fixation with natural death}), and clonal interference (Figure \ref{fig: fixation before CI}) leads to the failure of key results in evolutionary graph theory such as amplification of selection and the circulation theorem.
On the other hand, deriving evolutionary graph theory from ecologically motivated assumptions provided new insights. 
We were able to interpret evolutionary graph theory dynamics in terms of birth and death rates (Table \ref{tab: derived standard dynamics}), and show that natural death can prevent amplification of selection (Figure \ref{fig: comparison analytical}).


Representing network vertices as sites rather than individuals, which evolutionary graph theory assumes, is the key step we took and others often take \cite{schimit2019,broom2020} to develop a more general model.
This allowed us to incorporate intra-site dynamics and develop the network birth and death model (NBD). 
In NBD, intra-site competition is taken from Huang et al.~\cite{huang2015}, which provides a way to consider evolutionary games \citep{maynardsmith1982}, including multi-player social dilemmas \cite{broom2018}.
The NBD model showed that, more so than network structure, allowing intra-site competition significantly increased the success of an advantageous mutant, which increased further when natural death was disallowed (Figure \ref{fig: fixation with natural death}).
Furthermore, when intra-site competition was intensified by suppressing ecological dynamics in NBD to obtain evolutionary graph theory, the effect of network structure was more pronounced.
This means site capacity, which is determined by intra-site competition, is an additional variable that can be considered when investigating the effect of network structure.

In evolutionary graph theory, comparisons are often made between birth-death updating, death-birth updating and sometimes a combination of the two e.g.~\cite{zukewich2013,kaveh2015}.
By suppressing the ecological dynamics in NBD, we automatically obtained dynamics where birth-death and death-birth updating is combined (Equations \eqref{eq: rep rate} and \eqref{eqReplacementRate}).
The parameters in these dynamics are ecologically motivated (in terms of birth and death rates), and the birth-death or death-birth component can be muted.
These dynamics are therefore easier to understand and allow us to switch between different types of updating rules.
In these dynamics, increasing the natural death rate increases the effect of the death-birth component and decreases the effect of the birth-death component. 
By altering the natural death rate, we were able to find that the star network amplifies selection because Bd dynamics (no natural death) allows sites to act as sources, i.e.~net exporters of offspring.
On the other hand, we show numerically (Figure \ref{fig: comparison analytical}) that allowing natural death can prevent these sites from being sources, causing the star network to act as a suppressor.
This suggests that amplification of selection requires dynamics in which source sites can exist.
The effect of natural death in preventing amplification does indeed extend to other networks that are amplifiers under Bd dynamics \cite{tkadlec2020}, although it is not specified whether these networks have source sites.
Therefore, the question remains whether the existence of source sites is a requirement in general for networks that amplify selection.

Tkadlec et al.~\cite{tkadlec2019} constructed networks that increase the fixation probability of a mutant, but at a cost of higher fixation time.
As per our investigations, these networks would be more susceptible to clonal interference.
In the networks we considered, we found that the fixation probability decreases as the mutation rate increases when there is clonal interference (Figure \ref{fig: fixation before CI}).
Frean et al.~\citep{frean2013} suggested that networks with higher fixation time are more susceptible to clonal interference. They showed that the star network has a higher fixation time than the complete network.
This is consistent with our observations.
Clonal interference is therefore another element that can be considered in the context of amplifiers, especially in those networks with long fixation times.

Extending evolutionary graph theory by considering movement continues to be an active area of research e.g.~\cite{schimit2019,broom2020}.
We implemented movement on a network-structured population combining birth with movement such that offspring can be placed on a different connected site from their parent.
This means that the movement is local and dependent upon the network-structure of the population.
With the exception of bD dynamics, we were able to recover all other standard evolutionary graph theory dynamics (Table \ref{tab: derived standard dynamics}).
This shows that the localised movement in standard evolutionary graph theory dynamics is primarily based on this mechanism where movement is combined with birth, but also highlights that other options exist. 
We could implement different movement dynamics by adding an additional term to the infinitesimal dynamics (Equation \eqref{eq: generator}) to account for a change in state caused by the movement of individuals.
This includes movements that would enable bD dynamics to be recovered.
%
%
%
%

\section*{Acknowledgement}
\noindent
The study was supported by EPSRC  grants EP/N014499/1 and EP/T031727/1.

\appendix

\section{Generator Details}\label{app: Generator Details}
Here we provide details on the generator and hitting probability given in Section \ref{sec2}. 
Though not in the main text,  details of the hitting time are also provided. 

The infinitesimal generator describes how the expected values of functions of our model change in infinitesimal time intervals. For a function $f$ acting on the stochastic process $\Sigma(t)$, the infinitesimal generator, $\mathcal{L}$, is defined as \cite{oksendal2013}
\begin{equation}
\mathcal{L}f(y)=\frac{d}{dt}\mathbb{E}[f(\Sigma(t))]|_{t=0}=\lim\limits_{t \to 0} \frac{\mathbb{E}[f(\Sigma(t))]-f(y)}{t}.
\nonumber
\end{equation}

\subsection{Hitting Probability}
The hitting probability of a state $A \in S$ is the probability that the Markov process eventually reaches state $A$, given that it started in some state $i$. Let $T^A$ be the time when the Markov process first enters state $A$, then the hitting probability from an initial state $i$ is given by
\begin{equation}
h_A(i)=P(T^A < \infty|\Sigma(0)=i).
\nonumber
\end{equation}
Using the infinitesimal generator, we can find equations describing the hitting probability. From the definition of the generator, we have
\begin{equation}
\mathcal{L}h_A(i)=\frac{d}{dt}\mathbb{E}[h_A(\Sigma(t))|\Sigma(0)=i].
\nonumber
\end{equation}
Given that the Markov process starts in state $i$, the expected value of the hitting probability does not change with time, and therefore this derivative must be equal to zero, giving
\begin{equation}
\mathcal{L}h_A(i)=0. \nonumber
\end{equation}
If our initial state $i=A$, then the hitting probability is equal to $1$, so we have $h_A(A)=1$. 
In summary, the hitting probability is given by solving
\begin{linenomath*}
\begin{align}
\begin{cases}
	\mathcal{L}h_{A}(i)=0, \\
	h_{A}(A)=1.
\end{cases}
\end{align}
\end{linenomath*}

\subsection{Hitting Time}
The expected time until the Markov process reaches a state $A$ from an initial state $i$, is defined as
\begin{equation}
k_A(i)=\mathbb{E}[T^A |\Sigma(0)=i].
\nonumber
\end{equation}
From the definition of the generator, we have
\begin{equation}
\mathcal{L}k_A(i)=\frac{d}{dt}\mathbb{E}[k_A(\Sigma(t))|\Sigma(0)=i].
\nonumber
\end{equation}
The derivative can be calculated by
\begin{equation}
\frac{d}{dt}\mathbb{E}[k_A(\Sigma(t))|\Sigma(0)=i]=\lim\limits_{h \to 0}\frac{\mathbb{E}[k_A(\Sigma(t+h))|\Sigma(0)=i]-\mathbb{E}[k_A(\Sigma(t))|\Sigma(0)=i]}{h}
\nonumber
\end{equation}
Since both of the expectations on the right-hand side condition on $\Sigma(0)=i$, the expected hitting time from $0$ must be equal. The expected time from $t+h$ therefore has to be $h$ less than the expected time from $t$, so this becomes
\begin{equation}
\frac{d}{dt}\mathbb{E}[k_A(\Sigma(t))|\Sigma(0)=i]=\lim\limits_{h \to 0}\frac{-h}{h}=-1.
\nonumber
\end{equation}
Therefore, it must hold that
\begin{equation}
\mathcal{L}k_A(i)=-1.
\nonumber
\end{equation}
If our initial state $i=A$, then the expected hitting time is equal to $0$, so we have $k_A(A)=0$. 
To summarise, the expected hitting time is given by solving
\begin{linenomath*}
\begin{align}
\begin{cases}
	\mathcal{L}k_{A}(i)=-1, \\
	k_{A}(A)=0.
\end{cases}
\end{align}
\end{linenomath*}

\section{Hitting probability for modified dynamics}\label{app: Hitting probability}
Here we show how the hitting probability for modified dynamics, Equation \eqref{eq:stationary}, is obtained.
We have that
\begin{linenomath*}\begin{align*}
	0 =& \lim_{c\to \infty} \mathcal{L}_ch_{\set{A}}(\set{S})\\
	0 =&\lim_{c\to\infty} 
		\sum_{i\in \set{S}}\sum_{x\in\set{X}}[1-\mu(i)]B(c,i,x,\set{S})[h_{\set{A}}(\set{S}\cup\{(U_i,x)\})-h_{\set{A}}(\set{S})]\\
		&+ \sum_{i\in\set{S}}\sum_{x\in\set{X}}\mu(u)B(c,i,x,\set{S})\int_{\mathbb{R}^l}[h_{\set{A}}(\set{S}\cup\{(w,x)\}-h_{\set{A}}(\set{S})]M(U_i,w) dw\\
		&+ \sum_{i\in\set{S}}D(c,i,\set{S})[h_{\set{A}}(\set{S}\setminus\{i\})-h_{\set{A}}(\set{S})].
\end{align*}\end{linenomath*}
Let
\begin{linenomath*}\begin{align*}
\lambda_\set{S} = 
\lim_{c\to\infty}\sum_{i\in \set{S}}\sum_{x\in\set{X}}B(c,i,x,\set{S}) + D(c,i,\set{S}).
\end{align*}\end{linenomath*}
then rearranging gives
\begin{linenomath*}\begin{align*}
	h_{\set{A}}(\set{S}) =&\lim_{c\to\infty}\frac{1}{\lambda_\set{S}}
		\sum_{i\in \set{S}}\sum_{x\in\set{X}} 
		B(c,i,x,\set{S})\bigg([1-\mu(i)]h_{\set{A}}(\set{S}\cup\{(U_i,x)\})\\
		&+ \mu(i)\int_{\mathbb{R}^l}h_{\set{A}}(\set{S}\cup\{(w,x)\})M(U_i,w) dw\bigg)\\
		&+ D(c,i,\set{S})h_{\set{A}}(\set{S}\setminus\{i\}).
\end{align*}\end{linenomath*}
We assume that the population starts in a state $\set{S}$ where $|\set{S}|_x=1\ \forall x\in \set{X}$, we then have that 
\begin{linenomath*}\begin{align*}
	h_{\set{A}}(\set{S}) =& \frac{1}{\lambda_\set{S}}
		\sum_{i\in \set{S}}\sum_{x\in\set{X}}b(i,x,\set{S})\bigg([1-\mu(i)]h_{\set{A}}(\set{S}\cup\{(U_i,x)\})\\
		&+ \mu(i)\int_{\mathbb{R}^l}h_{\set{A}}(\set{S}\cup\{(w,x)\})M(U_i,w) dw\bigg)\\
		&+ d(i,\set{S})h_{\set{A}}(\set{S}\setminus\{i\})
\end{align*}\end{linenomath*}
since all sites have 1 individual.
From state $\set{S}$ we consider the following different states $\set{J}$ that the population can transition to.

\begin{enumerate}
\item For $\set{J}=\set{S}\cup\{(u,x)\}$ we have that
\begin{linenomath*}\begin{align*}
	h_{\set{A}}(\set{J}) =& \lim_{c\to\infty}\frac{1}{\lambda_\set{J}}
		\sum_{j\in \set{J}}\sum_{y\in\set{X}}[1-\mu(j)]B(c,j,y,\set{J})h_{\set{A}}(\set{J}\cup\{j\})\\
		&+ \mu(j)B(c,j,y,\set{J})\int_{\mathbb{R}^l}h_{\set{A}}(\set{J}\cup\{(w,y)\}M(U_j,w) dw\\
		&+ D(c,j,\set{J})h_{\set{A}}(\set{J}\setminus\{j\}).
\end{align*}\end{linenomath*}
The birth rate in this case is given by
\begin{linenomath*}\begin{align*}
	\lim_{c\to\infty}B(c,j,y,\set{J}) = \lim_{c\to\infty}c^{H_0[-|\set{J}_y|]}b(j,y,\set{J})=b(j,y,\set{J})
\end{align*}\end{linenomath*}
as $H_0[-|\set{J}|_y]=0\ \forall y\in\set{X}$ as there are no empty sites.
Similarly, the death in this case is given by
\begin{linenomath*}\begin{align*}
	\lim_{c\to\infty}D(c,j,\set{J})
	=\lim_{c\to\infty}c^{H_2[|\set{S}_{X_j}|]}d(j,\set{J})
	=\lim_{c\to\infty}c^{\delta_{X_j,x}}d(j,\set{J})
\end{align*}\end{linenomath*}		
as site $x$ is the only site with two individuals and $\delta_{m,n}$ is the Kronecker delta function.
This means that 
\begin{linenomath*}\begin{align*}
	\frac{\displaystyle\lim_{c\to\infty}B(c,j,y,\set{J})}{\lambda_\set{J}}=0\quad \forall j\in \set{J},y\in\set{X}
\end{align*}\end{linenomath*}
and
\begin{linenomath*}\begin{align*}
	\frac{\displaystyle\lim_{c\to\infty}	D(c,j,\set{J})}{\lambda_\set{J}}=
	\frac
	{\displaystyle \lim_{c\to\infty}c^{\delta_{X_j,x}}d(j,\set{J})}
	{\displaystyle \sum_{j\in\set{J}}\lim_{c\to\infty}c^{\delta_{X_j,x}}d(j,\set{J})}
	= 
	\frac
	{{\delta_{X_j,x}}d(j,\set{J})}
	{\displaystyle \sum_{j\in\set{J}_x}{d(j,\set{J})}}.
\end{align*}\end{linenomath*}
The hitting probability from state $\set{J}$ is then given by
\begin{linenomath*}\begin{align*}
	h_{\set{A}}(\set{J})= 
	\sum_{j\in\set{J}_x}
	\frac
	{d(j,\set{J})h_{\set{A}}(\set{J}\setminus\{j\})}
	{\displaystyle \sum_{k\in\set{J}_x}d(k,\set{J})}.
\end{align*}\end{linenomath*}
\item For $\set{J}=\set{S}\setminus\{i\}$ such that $i\in\set{S}$, by following a similar set of arguments as we have for case 1 we obtain the hitting probability from state $\set{J}$ as follows
\begin{linenomath*}\begin{align*}
	h_{\set{A}}(\set{J})=&
	\sum_{j\in\set{J}}
	\frac
	{\displaystyle b(j,X_i,\set{J})}
	{
		\displaystyle
		\sum_{j\in\set{J}}
		b(j,X_i,\set{J})
	}
	\bigg([1-\mu(j)]h_{\set{A}}(\set{J}\cup\{(U_j,X_i)\})\\
	& 
	+ \mu(j)\int_{\mathbb{R}^l}h_{\set{A}}(\set{J}\cup\{(w,X_i)\})M(U_j,w) dw\bigg).
\end{align*}\end{linenomath*}
\end{enumerate}

Substituting the hitting probability from $\set{J}$ for these two cases into the hitting probability from $\set{S}$ gives
\begin{linenomath*}\begin{align*}
	h_{\set{A}}(\set{S}) =& \frac{1}{\lambda_\set{S}}
		\sum_{i\in \set{S}}\sum_{x\in\set{X}}b(i,x,\set{S})
	\bigg(
		[1-\mu(i)]
		\frac
			{
				\displaystyle 
				\sum_{j\in\set{S}_x\cup\{(U_i,x)\}}
				\mkern-36mu 
				d(j,\set{S}\cup\{(U_i,x)\})
				h_{\set{A}}(\set{S}\cup\{(U_i,x)\}\setminus\{j\})
			}
			{
				\displaystyle 
				\sum_{j\in\set{S}_x\cup\{(U_i,x)\}}
				\mkern-36mu 
				d(j,\set{S}\cup\{(U_i,x)\})
			}
	\\
	& +\mu(i)\int_{\mathbb{R}^l}
		\frac
			{
				\displaystyle 
				\sum_{j\in\set{S}_x\cup\{(w,x)\}}				
				\mkern-36mu 
				d(j,\set{S}\cup\{(w,x)\})
				h_{\set{A}}(\set{S}\cup\{(w,x)\}\setminus\{j\})
			}
			{
				\displaystyle 
				\sum_{j\in\set{S}_x\cup\{(w,x)\}}				
				\mkern-36mu 
				d(j,\set{S}\cup\{(w,x)\})
			}
		M(U_i,w) dw\bigg)\\
		&+ d(i,\set{S})
	\sum_{j\in\set{S}\setminus\{i\}}
	\frac
	{\displaystyle b(j,X_i,\set{S}\setminus\{i\})}
	{
		\displaystyle
		\sum_{j\in\set{S}\setminus\{i\}}
		b(j,X_i,\set{S}\setminus\{i\})
	}
	\bigg([1-\mu(j)]h_{\set{A}}(\set{S}\setminus\{i\}\cup\{(U_j,X_i)\})\\ 
	&+ \mu(j)\int_{\mathbb{R}^l}h_{\set{A}}(\set{S}\setminus\{i\}\cup\{(w,X_i)\})M(U_j,w) dw\bigg)
\end{align*}\end{linenomath*}
This can be rewritten as follows
\begin{linenomath*}\begin{align*}
	h_{\set{A}}(\set{S}) =& \frac{1}{\lambda_\set{S}}
		\sum_{i\in \set{S}}
		\sum_{j\in\set{S}}
		\left(
		b(i,X_j,\set{S})
		\frac
			{
				\displaystyle 
				d(j,\set{J})
			}
			{
				\displaystyle 
				\sum_{k\in\set{J}_{X_j}}
				\mkern-6mu 
				d(k,\set{J})
			}
		+
		d(j,\set{S})
		\frac
		{\displaystyle b(i,X_j,\set{S}\setminus\{j\})}
		{
			\displaystyle
			\sum_{k\in\set{S}\setminus\{j\}}
			b(k,X_i,\set{S}\setminus\{j\})
		}
		\right)
	\\
	&\times
		[1-\mu(i)]		
		h_{\set{A}}(\set{J}\setminus\{j\})
	\\
	&+
	\int_{\mathbb{R}^l}
	\left(
		b(i,X_j,\set{S})
		\frac
			{
				\displaystyle 
				d(j,\set{K})
			}
			{
				\displaystyle 
				\sum_{k\in\set{K}_{X_j}}
				\mkern-6mu 
				d(k,\set{K})
			}
		+
		d(j,\set{S})
		\frac
		{\displaystyle b(i,X_j,\set{S}\setminus\{j\})}
		{
			\displaystyle
			\sum_{k\in\set{S}\setminus\{j\}}
			b(k,X_i,\set{S}\setminus\{j\})
		}
		\right)
		\\
		&\times	
		\mu(i)		
		h_{\set{A}}(\set{K}\setminus\{j\})M(U_i,w) dw
\end{align*}\end{linenomath*}
where $\set{J}=\set{S}\cup\{(U_i,X_j)\})$ and $\set{K}=\set{S}\cup\{(w,X_j)\})$.
This can then be further simplified by writing
\begin{linenomath*}\begin{align}
	h_{\set{A}}(\set{S}) =& \frac{1}{\lambda_\set{S}}
		\sum_{i\in \set{S}}
		\sum_{j\in\set{S}}
		r(i,j,U_i,\set{S})
		[1-\mu(i)]		
		h_{\set{A}}(\set{J}\setminus\{j\})
	\nonumber\\
	&+
	\int_{\mathbb{R}^l}
		r(i,j,w,\set{S})	
		\mu(i)		
		h_{\set{A}}(\set{K}\setminus\{j\})M(U_i,w) dw
\label{eq: stationary with replacement}
\end{align}\end{linenomath*}
where $r(i,j,u,\set{S})$ is the rate at which the offspring of $I_i$ replaces $I_j$ given that the offspring has trait $u$.

\section{Infinitesimal generator and hitting probability for evolutionary graph theory}\label{app: Infinitesimal generator}
Here we provide the definition of the infinitesimal generator used and the hitting probability obtained for evolutionary graph theory mentioned in Section \ref{sec: app 1}.
The infinitesimal generator for evolutionary graph theory is defined as follows 
\begin{linenomath*}\begin{align}
	\mathcal{L}^\text{EGT}\phi(\set{S}) =&\sum_{i\in \set{S}}
		\sum_{j\in\set{S}}
		[1-\mu(i)]		
		R(i,j,U_i,\set{S})
	\nonumber\\
	&\times
		[\phi(\set{S}\cup\{(U_i,X_j)\}\setminus\{j\})-\phi(\set{S})]
	\nonumber\\
	&+
	\mu(i)
	\int_{\mathbb{R}^l}
		R(i,j,w,\set{S})
	\nonumber\\
	&\times	
		[\phi(\set{S}\cup\{(w,X_j)\}\setminus\{j\})-\phi(\set{S})]
		M(U_i,w) dw
\end{align}\end{linenomath*}
where $R$ is the replacement rate in evolutionary graph theory dynamics. 
Note that $R$ is a function of $W$ but has been dropped for brevity.
This generator with continuous mutations has not been considered before but it allows direct comparisons between $h^\text{NBD}$ and $h^\text{EGT}$.
In particular, solving Equation \eqref{eq: hitting prob} with $\mathcal{L}^\text{EGT}$ gives the hitting probability in evolutionary graph theory,
\begin{linenomath*}\begin{align}
	h_{\set{A}}^\text{EGT}(\set{S}) =&\frac{1}{\lambda_\set{S}}\sum_{i\in \set{S}}
		\sum_{j\in\set{S}}
		R(i,j,U_i,\set{S})
		[1-\mu(i)]
	\nonumber\\
		&\times		
		h_{\set{A}}^\text{EGT}(\set{S}\cup\{(U_i,X_j)\}\setminus\{j\})
	\nonumber\\
	&+
	\int_{\mathbb{R}^l}
		R(i,j,w,\set{S})	
		\mu(i)
	\nonumber\\
	&\times		
		h_{\set{A}}^\text{EGT}(\set{S}\cup\{(w,X_j)\}\setminus\{j\})M(U_i,w) dw,
\label{eq: hitting egt}
\end{align}\end{linenomath*}
where $\lambda_{\set{S}}$ is the rate of leaving state $\set{S}$. 
Note that its form is similar that of $h^\text{NBD}$.

\section{Deriving standard evolutionary graph theory dynamics}\label{app: Deriving standard}
Here we show how we derive standard evolutionary graph theory dynamics from NBD, see Section \ref{sec: app 1}. 

We need to show that 
\begin{align*}
     h^\text{NBD}_{\set{A}}(\set{S})=h^\text{EGT}_{\set{A}}(\set{S}).
\end{align*}
We start by observing that for all the standard evolutionary graph theory dynamics, the replacement rate satisfies
\begin{linenomath*}\begin{align}
	R(i,j,u,\set{S})=R(i,j,v,\set{S})\ \forall u,v\in\set{U}
	\label{eq: condition required}
\end{align}\end{linenomath*}
and therefore
\begin{linenomath*}\begin{align*}
	\lambda_\set{S} 
		&=
		\sum_{i\in \set{S}}
		\sum_{j\in\set{S}}
		R(i,j,U_i,\set{S})
		[1-\mu(i)]
		+ 
		\int_{\mathbb{R}^l}
		R(i,j,w,\set{S})	
		\mu(i)		
		M(U_i,w) dw
	\\
		&=
		\sum_{i\in \set{S}}
		\sum_{j\in\set{S}}
		R(i,j,\set{S}),	
\end{align*}\end{linenomath*}
where $R(i,j,\set{S})$ is the replacement rate with the type of the offspring dropped.
Furthermore, for all the standard evolutionary graph theory dynamics the following also holds
\begin{linenomath*}\begin{align*}
	\lambda_\set{S}=
		\sum_{i\in \set{S}}
		\sum_{j\in\set{S}}
		R(i,j,\set{S})
		= 1
\end{align*}\end{linenomath*}		
since the replacement rates are defined as probabilities.
We then require that the replacement rate $r$ for NBD have the same property as in Equation \eqref{eq: condition required}, that is,
\begin{linenomath*}\begin{align}
	r(i,j,u,\set{S})=r(i,j,v,\set{S})\ \forall u,v\in\set{U},
	\label{eq: condition required r}
\end{align}\end{linenomath*}
we can therefore use $r(i,j,\set{S})$ as the offspring type can be dropped, and 
\begin{linenomath*}\begin{align*}
	R(i,j,\set{S}) = 
	\frac{r(i,j,\set{S})}
	{
		\displaystyle
		\sum_{n\in \set{S}}
		\sum_{k\in\set{S}}
		r(n,k,\set{S})
	}.
\end{align*}\end{linenomath*}
This ensures that the hitting probability is identical for both types of dynamics.
Recall that the replacement rate for NBD is given by
\begin{linenomath*}\begin{align*}
	r(i,j,u,\set{S})=
	s\beta_{U_i}W_{X_i,X_j}
	\frac
	{
		\delta_{U_j} + \gamma_{U_j,u}
	}
	{
		\delta_{U_j} + \gamma_{U_j,u} + \delta_{u} + \gamma_{u,U_j}
	}
	+
	\delta_{U_j}
	\frac
	{
		\beta_{U_i}W_{X_i,X_j}
	}
	{
		\displaystyle
		\sum_{k\in\set{S}\setminus\{j\}} \beta_{U_k}W_{X_k,X_j}
	}.
\end{align*}\end{linenomath*}
We can now consider which of the standard evolutionary graph theory dynamics we can obtain from these dynamics.

\paragraph{LB dynamics}
Setting $\delta_{u}=0$ and $\gamma_{u,v}=\gamma_{v,u} \ \forall u,v\in\set{U}$ satisfies Equation \eqref{eq: condition required r} and gives
\begin{linenomath*}\begin{align*}
\frac{r(i,j,\set{S})}
	{
		\displaystyle
		\sum_{n\in \set{S}}
		\sum_{k\in\set{S}}
		r(n,k,\set{S})
	}
=
\frac{\beta_{U_i}W_{X_i,X_j}}
	{
		\displaystyle
		\sum_{n\in \set{S}}
		\sum_{k\in\set{S}}
		\beta_{U_n}W_{X_n,X_k}
	}
\end{align*}\end{linenomath*}
which is identical to the LB dynamics when $\beta_u=F_u\ \forall u\in\set{U}$. 

\paragraph{Bd dynamics}
Doing the same as with the derivation of LB dynamics, but setting $W\mat{1}=\mat{1}$ gives
\begin{linenomath*}\begin{align*}
\frac{r(i,j,\set{S})}
	{
		\displaystyle
		\sum_{n\in \set{S}}
		\sum_{k\in\set{S}}
		r(n,k,\set{S})
	}
=
\frac{\beta_{U_i}}
	{
		\displaystyle
		\sum_{n\in \set{S}}
		\beta_{U_n}
	}
	W_{X_i,X_j}
\end{align*}\end{linenomath*}
which is identical to the Bd dynamics when $\beta_u=F_u\ \forall u\in\set{U}$. 

\paragraph{Db dynamics}
Setting $s=0$ and $\beta_{u}=\beta_{v} \ \forall u,v\in\set{U}$ satisfies Equation \eqref{eq: condition required r} and gives
\begin{linenomath*}\begin{align*}
\frac{r(i,j,\set{S})}
	{
		\displaystyle
		\sum_{n\in \set{S}}
		\sum_{k\in\set{S}}
		r(n,k,\set{S})
	}
=\frac{
	\displaystyle
	\delta_{U_j}
	\frac
	{
		\displaystyle
		W_{X_i,X_j}
	}
	{
		\displaystyle
		\sum_{k\in\set{S}\setminus\{j\}} W_{X_k,X_j}
	}
}{
	\displaystyle
	\sum_{n\in \set{S}}
	\sum_{k\in\set{S}\setminus\{n\}}
	\delta_{U_n}
	\frac
	{
		\displaystyle
		W_{X_k,X_n}
	}
	{
		\displaystyle
		\sum_{m\in\set{S}\setminus\{n\}} W_{X_m,X_n}
	}
}
=\frac{
	\delta_{U_j}
	}
	{
	\displaystyle
	\sum_{n\in \set{S}}
	\delta_{U_n}
	}
	\frac
	{
		\displaystyle
		W_{X_i,X_j}
	}
	{
		\displaystyle
		\sum_{k\in\set{S}\setminus\{j\}} W_{X_k,X_j}
	}
\end{align*}\end{linenomath*}	
which is identical to Db dynamics when $\delta_u=1/F_u\ \forall u\in\set{U}$.

\paragraph{dB Dynamics}
Setting $s=0$ and $\delta_{u}=\delta_{v} \ \forall u,v\in\set{U}$ satisfies Equation \eqref{eq: condition required r} and gives
\begin{linenomath*}\begin{align*}
\frac{r(i,j,\set{S})}
	{
		\displaystyle
		\sum_{n\in \set{S}}
		\sum_{k\in\set{S}}
		r(n,k,\set{S})
	}
=\frac{
	\displaystyle
	\delta_{U_j}
	\frac
	{
		\displaystyle
		\beta_{U_i}W_{X_i,X_j}
	}
	{
		\displaystyle
		\sum_{k\in\set{S}\setminus\{j\}} \beta_{U_k}W_{X_k,X_j}
	}
}{
	\displaystyle
	\sum_{n\in \set{S}}
	\sum_{k\in\set{S}\setminus\{n\}}
	\delta_{U_n}
	\frac
	{
		\displaystyle
		\beta_{U_k}W_{X_k,X_n}
	}
	{
		\displaystyle
		\sum_{m\in\set{S}\setminus\{n\}} \beta_{U_m}W_{X_m,X_n}
	}
}
=
\frac{1}{N}
\frac
	{
		\displaystyle
		\beta_{U_i}W_{X_i,X_j}
	}
	{
		\displaystyle
		\sum_{k\in\set{S}\setminus\{j\}} \beta_{U_k}W_{X_k,X_j}
	}
\end{align*}\end{linenomath*}	
which is identical to dB dynamics when $\beta_u=F_u\ \forall u\in\set{U}$. 			

\paragraph{LD Dynamics}
These require the competition rate $\gamma_{u,v}$ and therefore \eqref{eq: condition required r} cannot be satisfied. 
However, we can bypass this condition by assuming there is no mutation and that there are only two types, i.e. $|\set{U}|=2$. 
Note that excluding transitions to the same state will not affect $h_{\set{A}}$ so if we discount transitions to the same state, we would require that
\begin{linenomath*}\begin{align}
		\frac{
			R(i,j,\set{S})
		}
		{
			\displaystyle
			\sum_{n\in \set{S}}
			\sum_{\substack{k\in\set{S}\\U_k\ne U_n}}
			R(n,k,\set{S})		
		} 
		=
		\frac{
			r(i,j,\set{S})
		}
		{
			\displaystyle
			\sum_{n\in \set{S}}
			\sum_{\substack{k\in\set{S}\\U_k\ne U_n}}
			r(n,k,\set{S})		
		} 
		\quad \text{for } U_j\ne U_i.
\label{eq: no self trans}
\end{align}\end{linenomath*}
Setting $\delta_{u}=0$ and $\beta_{u}=\beta_{v} \ \forall u\in\set{U}$ simplifies the RHS of Equation \eqref{eq: no self trans} to
\begin{linenomath*}\begin{align*}
\frac
{
	\displaystyle
	W_{X_i,X_j}
	\frac{\gamma_{U_j,U_i}}{\gamma_{U_j,U_i}+\gamma_{U_i,U_j}}}
{
	\displaystyle
	\sum_{n\in \set{S}}
	\sum_{\substack{k\in\set{S}\\U_k\ne U_n}}
	W_{X_n,X_k}\frac{\gamma_{U_k,U_n}}{\gamma_{U_k,U_n}+\gamma_{U_n,U_k}}
}
=
\frac
{
	\displaystyle
	W_{X_i,X_j}
	\gamma_{U_j,U_i}
}
{
	\displaystyle
	\sum_{n\in \set{S}}
	\sum_{\substack{k\in\set{S}\\U_k\ne U_n}}
	W_{X_n,X_k}
	\gamma_{U_k,U_n}
}
\end{align*}\end{linenomath*}
which for $\gamma_{u,v}=1/F_u$ when $u\ne v\ \forall u,v\in\set{U}$  is equivalent to the LHS of Equation \eqref{eq: no self trans} when using LD dynamics.

\section{Derivation of bias ($r_k$) in complete network for NBD}\label{app: Derivation of bias}
Here we show how the bias given in Equation \eqref{eq: bias rk} is obtained.
We start by defining the replacement rate in the complete network.
For ecological dynamics are suppressed, we only need to consider population states with one individual on each site.
The position of residents and mutants does not matter in these states due to site homogeneity.
Therefore, states with the same number of mutants $k$ (which means there are $N-k$ residents) are lumped together and referred to by this number. 
We are interested in the rate at which the system transitions from some state $k$ to a state with an additional type $u$ individual; i.e. with $k-(-1)^u$ mutants. The replacement rate for such a transition is denoted $q_{k,u}$. For NBD, this is given by 
\begin{linenomath*}\begin{align}
q_{k,u}{}
= 
	k(N-k)
	\left(
	s\beta_{u}w
	\frac{
		\delta_{1-u}+\gamma_{1-u,u}
	}{
		\der{0}{}  + \der{1}{}+ \gamma_{0,1}{} + \gamma_{1,0}{}
	}
	+
	\delta_{1-u}
	\frac{
		\beta_{u}
	}{
		\beta_0(N-k-u) + \beta_1(k-1+u)
	}
	\right)
.
\nonumber
\end{align}\end{linenomath*}
The bias is then given by
\begin{align*}
    r_k = \frac{q_{k,0}}{q_{k,1}}.
\end{align*} 

\section{Showing strict order in bias for complete network}\label{app: Showing strict}
Here we want to show that Equation \eqref{eq:bias order},
$
	r_k^\text{(i)} > r_k^\text{(ii)} > r_k^\text{(iii)},
$
holds for the complete network. 
From Table \ref{tab: dynamics combs} we have
\begin{align*}
	&r_k^\text{(i)} = 
	\frac
	{\beta_0}
	{\beta_1}
	\frac
	{(N-k-1)\beta_0+k\beta_1}
	{(N-k)\beta_0+(k-1)\beta_1}, \\
	&r_k^\text{(ii)} = 
	\frac
	{\delta\frac{\beta_0}{(N-k)\beta_0+(k-1)\beta_1}+\beta_0w\frac{1}{2}}
	{\delta\frac{\beta_1}{(N-k-1)\beta_0+k\beta_1}+\beta_1w\frac{1}{2}}, \\ 
	&r_k^\text{(iii)} =\frac{ \beta_0}{\beta_1}.
\end{align*}
The denominator in all three cases is strictly positive because we are assuming that \mbox{$\beta_0,\beta_1,\delta,N,k,w$} are strictly positive, and that $k \in \{1,\ldots,N-1\}$. 
Let 
\begin{align*}
	x_k = (N-k-1)\beta_0+k\beta_1 \quad\text{and}\quad y_k = (N-k)\beta_0+(k-1)\beta_1.
\end{align*}
We then have that
\begin{align*}
		r_k^\text{(i)} 
		> 
		r_k^\text{(ii)}
	&\Leftrightarrow
		\frac
		{\beta_0}
		{\beta_1}
		\frac
		{x_k}
		{y_k}
	> 
		\frac
		{\delta\frac{\beta_0}{y_k}+\beta_0w\frac{1}{2}}
		{\delta\frac{\beta_1}{x_k}+\beta_1w\frac{1}{2}}
	\\
	&\Leftrightarrow
		\beta_0\beta_1\left(\delta + \frac{wx_k}{2}\right)
	> 
		\beta_0\beta_1\left(\delta+\frac{wy_k}{2}\right)
	\\
	&\Leftrightarrow
		x_k
	> 
		y_k
	\\
	&\Leftrightarrow
	\beta_1 > \beta_0. 
\end{align*}
Similarly, we have that
\begin{align*}
	r_k^\text{(ii)}  
	>
	r_k^\text{(iii)}
	&\Leftrightarrow
	\frac
	{\delta\frac{\beta_0}{y_k}+\beta_0w\frac{1}{2}}
	{\delta\frac{\beta_1}{x_k}+\beta_1w\frac{1}{2}}
	>
	\frac{ \beta_0}{\beta_1}
	\\
	&\Leftrightarrow
	\beta_0\beta_1\left({\frac{\delta}{y_k}+w\frac{1}{2}}\right)
	>
	\beta_0\beta_1\left(\frac{\delta}{x_k}+w\frac{1}{2}\right)
	\\
	&\Leftrightarrow
		x_k
	> 
		y_k
	\\
	&\Leftrightarrow
	\beta_1 > \beta_0.
\end{align*}
Since $r_k^\text{(i)}> r_k^\text{(ii)}$ and $r_k^\text{(ii)} > r_k^\text{(iii)}$, we have that $r_k^\text{(i)}> r_k^\text{(iii)}$.
Therefore, Equation \eqref{eq:bias order} holds when $\beta_1>\beta_0$, which we have assumed is true for an advantageous mutant invading a resident population.


\section{Average fixation probability in star network} \label{app: Average fixation}
Here we show how to calculate the average fixation probability in the star network, where it is used in Section \ref{sec512}.
In the star network, we only consider population states with one individual on each site as ecological dynamics are suppressed.
In such states, the position of residents and mutants present on leaves does not matter, since the leaves are identical.
The population state is then given by $(u,k)$ where $u$ is the centre individual's type and $k$ is the number of mutants on leaves ($N-1-k$ is the number of residents on leaves).
Since $u\in\{0,1\}$, in state $(u,k)$ there are $1-u \le k \le N-1-u$ mutants on the leaves provided that there is at least one mutant and resident in the population. 
Let $r(u,k,u',k')$ be the rate of transitioning from state $(u,k)$ to $(u',k')$. 
We only need to consider the two transitions where a change in state occurs.
First, for NBD a type $u$ centre can replace a type $1-u$ leaf with rate
\begin{linenomath*}\begin{align}
r(u,k,u,k-(-1)^u)&=
		k^{1-u}(N-1-k)^{u}
		\left(
		sw_c\beta_u
		\frac{
			\delta_{1-u}+\gamma_{1-u,u}
		}{
			\delta_0+\delta_1+\gamma_{0,1}+\gamma_{1,0}
		}
	+
		\delta_{1-u}
		\right).
\end{align}\end{linenomath*}
Second, a type $u$ centre is replaced by a type $1-u$ leaf with rate
\begin{linenomath*}\begin{align}
r(u,k,1-u,k)&=
		k^{1-u}(N-1-k)^{u}
		\left(
		sw_l\beta_{1-u}
		\frac{
			\delta_u+ \gamma_{u,1-u}
		}{
			\delta_0+\delta_1+\gamma_{0,1}+\gamma_{1,0}
		}
	+
		\delta_u
		\frac{
			\beta_{1-u}
		}{
			(N-1-k)\beta_0 + k\beta_{1}
		} 
		\right).
\end{align}\end{linenomath*}
Using these rates, we can calculate the average fixation probability. 
In Hadjichrysanthou et al \cite{hadjichrysanthou2011}, the average fixation probability is given by 
\begin{linenomath*}\begin{align*}
	\bar{\rho}_\text{star} = \frac{\displaystyle \rho_{\text{star(centre)}} + (N-1)\rho_{\text{star(leaf)}}}{N}.
\end{align*}\end{linenomath*}
Here, $\rho_{\text{star(centre)}}$ is the fixation probability of a mutant starting in the centre and $\rho_{\text{star(leaf)}}$ is the fixation probability of a mutant starting in a leaf. 
They are given by
\begin{linenomath*}\begin{align}
	\rho_{\text{star(centre)}}=\frac{p(1,0,1,1)}{A(1,N-1)} \quad \text{and}\quad 
	\rho_{\text{star(leaf)}}=\frac{p(0,1,1,1)}{A(1,N-1)}
\end{align}\end{linenomath*}		
where 
\begin{linenomath*}\begin{align*}
	A(l,m) = 1+\sum_{j=l}^{m-1} p(1,j,0,j) \prod_{k=l}^j \frac{p(0,k,0,k-1)}{p(1,k,1,k+1)}.
\end{align*}\end{linenomath*}
and
\begin{linenomath*}\begin{align*}
	p(u,k,u,k-(-1)^u)=\frac{r(u,k,u,k-(-1)^u)}{r(u,k,u,k-(-1)^u) + r(u,k,1-u,k) }, \\
	p(u,k,1-u,k)=\frac{r(u,k,1-u,k)}{r(u,k,u,k-(-1)^u) + r(u,k,1-u,k) }.
\end{align*}\end{linenomath*}
For the cases in Table \ref{tab: dynamics compared}, $\bar{\rho}_\text{star}$ is given by:\\
\begin{linenomath*}
\begin{align*}
\bar{\rho}_\text{star}^\text{(i)}=
\frac{
	\frac{1}{N}
	\frac{
		N-1	
	}{
		N 
	}
	+
	\frac{N-1}{N}
	\frac{
		\beta_{1}
	}{
		(N-2)\beta_0 + 2\beta_{1}
	}	
}
{
	\displaystyle
	1+
	\sum_{j=1}^{N-2}
	\frac{
		\beta_{0}
	}{
		(N-j)\beta_0 + j\beta_{1}
	}
	\prod_{k=1}^j
	\frac
	{
		(N-k)\beta_0 + k\beta_{1}
	}{
		(N-1-k)\beta_0 + (k+1)\beta_{1}
	}
},
\end{align*}
\end{linenomath*}
\begin{linenomath*}
\begin{align*}
\bar{\rho}_\text{star}^\text{(ii)}=
\frac{
	\frac{1}{N}
	\frac{
		sw_c\beta_1/2 +\delta	
	}{
		s(w_c\beta_1 +w_l\beta_0)/2 +\delta \frac{N}{N-1} 
	}
	+
	\frac{N-1}{N}
	\frac{
		sw_l\beta_{1}/2 + \delta\frac{\beta_{1}}{(N-2)\beta_0 + \beta_{1}}	
	}{
		s(w_l\beta_{1} + w_c\beta_0)/2 + \delta\frac{(N-2)\beta_0 + 2\beta_{1}}{(N-2)\beta_0 + \beta_{1}}
	}	
}
{
	1+
	\sum_{j=1}^{N-2}
	\frac{
		sw_l\beta_{0}/2	+ \delta \frac{\beta_{0}}{(N-1-j)\beta_0 + j\beta_{1}}
	}{
		s(w_l\beta_{0}+ w_c\beta_1)/2	+ \delta \frac{(N-j)\beta_0 + j\beta_{1}}{(N-1-j)\beta_0 + j\beta_{1}}
	}
	\prod_{k=1}^j
	\frac
	{
		sw_c\beta_0/2 +\delta
	}{
		sw_c\beta_1/2 +\delta
	}
	\frac
	{
		s(w_c\beta_1 + w_l\beta_{0})/2	+ \delta \frac{(N-k)\beta_0 + k\beta_{1}}{(N-1-k)\beta_0 + k\beta_{1}}		
	}{
		s(w_c\beta_0+ w_l\beta_{1})/2 + \delta\frac{(N-1-k)\beta_0 + (k+1)\beta_{1}}{(N-1-k)\beta_0 + k\beta_{1}}
	}
},
\end{align*}
\end{linenomath*}
\begin{linenomath*}
\begin{align*}
\bar{\rho}_\text{star}^\text{(iii)}=
\frac{
	\frac{1}{N}
	\frac{
		w_c\beta_1 	
	}{
		w_c\beta_1 +w_l\beta_0
	}
	+
	\frac{N-1}{N}
	\frac{
		w_l\beta_1 
	}{
		w_l\beta_{1} + w_c\beta_0
	}	
}
{
	\displaystyle
	1+
	\sum_{j=1}^{N-2}
	\frac{
		w_l\beta_{0}
	}{
		w_l\beta_{0}+ w_c\beta_1
	}
	\left(
	\frac
	{
		\beta_0
	}{
		\beta_1
	}
	\frac
	{
		w_c\beta_1 + w_l\beta_{0}
	}{
		w_c\beta_0+ w_l\beta_{1}
	}
	\right)^j
}.
\end{align*}
\end{linenomath*}

\section{Proof for star network}\label{app: Proof for star}
Here  we  want  to  show  that  Equation  \eqref{eq:bias order} holds for the star network when $N\to\infty$. 
In case (i) we have that
\begin{linenomath*}
\begin{align*}
	\lim_{N\to\infty} \bar{\rho}_\text{star}^\text{(i)}=0.
\end{align*}		 
\end{linenomath*}
In case (ii) we have that
\begin{linenomath*}
\begin{align*}
\lim_{N\to\infty} \bar{\rho}_\text{star}^\text{(ii)} = 
\frac{
	\frac{
		w_l\beta_{1} 
	}{
		w_l\beta_{1} + w_c\beta_0 + 2\delta
	}	
}
{
	\displaystyle
	1+
	\frac{
		w_l\beta_{0}	
	}{
		w_l\beta_{0}+ w_c\beta_1	+ 2\delta 
	}
	\sum_{j=1}^{\infty}
	\left(
		\frac
		{
			w_c\beta_0 +2\delta
		}{
			w_c\beta_1 +2\delta
		}
		\frac
		{
			w_c\beta_1 + w_l\beta_{0}	+ 2\delta
		}{
			w_c\beta_0+ w_l\beta_{1} + 2\delta
		}
	\right)^j
}.
\end{align*}
\end{linenomath*}
The denominator in this case converges to 
\begin{linenomath*}
\begin{align*}
	1 + \frac{ar}{1-r}
\end{align*}
\end{linenomath*}
where
\begin{linenomath*}
\begin{align*}
	a &= \frac{
		w_l\beta_{0}	
	}{
		w_l\beta_{0}+ w_c\beta_1	+ 2\delta 
	},\\
	r & =
		\frac
		{
			w_c\beta_0 +2\delta
		}{
			w_c\beta_1 +2\delta
		}
		\frac
		{
			w_c\beta_1 + w_l\beta_{0}	+ 2\delta
		}{
			w_c\beta_0+ w_l\beta_{1} + 2\delta
		}.
\end{align*}
\end{linenomath*}
Let 
\begin{linenomath*}
\begin{align*}
	x = \frac{
		w_l\beta_{1} 
	}{
		w_l\beta_{1} + w_c\beta_0 + 2\delta
	},
\end{align*}
\end{linenomath*}
we therefore have that 
\begin{linenomath*}
\begin{align*}
	\lim_{N\to\infty} \bar{\rho}_\text{star}^\text{(ii)} = \frac{x(1-r)}{1+r(a-1)}.
\end{align*}		
\end{linenomath*}
In case (iii) we have that 
\begin{linenomath*}
\begin{align*}
	\lim_{N\to\infty} \bar{\rho}_\text{star}^\text{(iii)} = \frac{x_{\delta=0}(1-r_{\delta=0})}{1+r_{\delta=0}(a_{\delta=0}-1)}
\end{align*}
\end{linenomath*}
where $x_{\delta=0},a_{\delta=0},r_{\delta=0}$ are $x,a,r$ with $\delta=0$. 
We have that 
\begin{linenomath*}
\begin{align*}
	\lim_{N\to\infty} \bar{\rho}_\text{star}^\text{(ii)} <
	\lim_{N\to\infty} \bar{\rho}_\text{star}^\text{(iii)} 
\end{align*}		
\end{linenomath*}
if $r>r_{\delta=0},a<a_{\delta=0},x<x_{\delta=0}$, which is indeed the case since $\beta_1>\beta_0$ and $\delta>0$ (in case (ii)). 
This therefore gives
\begin{linenomath*}
\begin{align*}
	\lim_{N\to\infty} \bar{\rho}_\text{star}^\text{(i)} <
	\lim_{N\to\infty} \bar{\rho}_\text{star}^\text{(ii)} <
	\lim_{N\to\infty} \bar{\rho}_\text{star}^\text{(iii)} 
\end{align*}
\end{linenomath*}
as required.

\section{Simulation Details}\label{app: Simulation Details}
The Gillespie alogrithm \cite{gillespie1976,gillespie1977} is used to simulate the evolutionary process described by the infinitesimal generator (Equation \eqref{eq: generator}),
\begin{linenomath*}
\begin{align*}
	\mathcal{L}\phi(\set{S}) &= 
		\sum_{i\in \set{S}}
		\sum_{n\in\set{X}}
		[1-\mu(i)]b(i,n,\set{S})[\phi(\set{S}\cup\{(U_i,n)\})-\phi(\set{S})]
		\nonumber\\
		&+ \sum_{i\in\set{S}}
		\sum_{n\in\set{X}}
		\mu(i)b(i,n,\set{S})\int_{\mathbb{R}^l}[\phi(\set{S}\cup\{(w,n)\}-\phi(\set{S})]M(U_i,w) dw
		\nonumber\\
		&+ 
		\sum_{i\in\set{S}}d(i,\set{S})[\phi(\set{S}\setminus\{i\})-\phi(\set{S})].
\end{align*}
\end{linenomath*}
Let $T(k)$ be the time and $S(k)$ be the state of the population after $k$ events have taken place. 
The following steps are followed  for the simulation.
\begin{enumerate}
\item Determine the time, $T(k+1)$, when a new event happens as follows,
\begin{align*}
	T(k+1) = T(k) - \frac{\ln(\mathcal{U}(0,1))}{\lambda_k}
\end{align*}
where 
\begin{align*}
\lambda_k = 
	\sum_{i\in S(k)}\sum_{x\in\set{X}}b(i,x,S(k)) + d(i,S(k)).
\end{align*}
and $\mathcal{U}(0,1)$ is a random number uniformly distributed in the range $(0,1)$. 
\item Determine the state, $S(k+1)$, when a new event takes place:
\begin{itemize}
\item Birth without mutation: $I_i$ gives birth to an offspring of the same type onto site $n$ with probability
	\begin{align*}
	[1-\mu(i)]\frac{b(i,n,S(k))}{\lambda_k},
	\end{align*}
	then 
	\begin{align*}
		S(k+1) = S(k) \cup\{(U_i,n)\}.
	\end{align*}
\item Birth with mutation: $I_i$ gives birth to an offspring of type $w$ onto site $n$ with probability
	\begin{align*}
	\mu(i)M(U_i,w)\frac{b(i,n,S(k))}{\lambda_k},
	\end{align*}
	then 
	\begin{align*}
		S(k+1) = S(k) \cup\{(w,n)\}.
	\end{align*}
\item Death: $I_i$ dies with probability 
\begin{align*}
	\frac{d(i,S(k))}{\lambda_k},
\end{align*}
then 
\begin{align*}
	S(k+1) = S(k) \setminus\{i\}.
\end{align*}

\end{itemize}

\item Repeat step 1 and 2 as necessary.
\end{enumerate}

To solve the hitting probability (Equation \eqref{eq: fixation}), 
\begin{linenomath*}
\begin{align*}
	\begin{cases}
		\mathcal{L}_c\rho(\set{S})=0\quad &\set{S}\notin \set{M}\cup\set{R},\\
		\rho(\set{S})=0 \quad &\set{S}\in \set{R},\\
		\rho(\set{S})=1\quad &\set{S} \in \set{M},
	\end{cases}
\end{align*}
\end{linenomath*}
we set $T(0)=0$ and $S(0) = \set{S}$ such that $\set{S}\notin \set{M}\cup\set{R}$, and then repeat steps 1 and 2 in the above alogrithm until we hit a state in $\set{M}$ or $\set{R}$. 
If we run $N_\text{sim}$ simulations, out of which $N_\text{mut}$ hit a state in $\set{M}$, then the hitting probability is given by
\begin{align*}
	\rho(\set{S})= \frac{N_\text{mut}}{N_\text{sim}}.
\end{align*}

\bibliographystyle{agsm}
\bibliography{Project1}

\end{document}